\documentclass[12pt,preprint]{aastex}

\newcommand{\bm}[1]{\mbox{\boldmath $#1$}}
\usepackage{graphicx}
\usepackage{psfig}

\slugcomment{}

\title{Vertical Structure of Gas Pressure-Dominated Accretion Disks
with Local Dissipation of Turbulence and Radiative Transport}

\shortauthors{Hirose et al.}

\begin{document}

\shorttitle{Vertical Structure of Gas-Dominated Accretion Disks}

\author{Shigenobu Hirose and Julian H. Krolik}
\affil{Department of Physics and Astronomy, Johns Hopkins University, 
    Baltimore, MD 21218}


\and

\author{James M. Stone}
\affil{Department of Astrophysical Sciences, 
Princeton University, Princeton, NJ 08544}

\begin{abstract}

We calculate the vertical structure of a local patch of an
accretion disk in which heating by dissipation of MRI-driven
MHD turbulence is balanced by radiative cooling.
Heating, radiative transport, and cooling are computed
self-consistently with the structure by solving the equations of
radiation MHD in the shearing-box approximation.  Using
a fully 3-d and energy-conserving code, we compute the structure
of this disk segment over a span of more than five cooling times.
After a brief relaxation period, a statistically steady-state develops.
Measuring height above the midplane in units of the scale-height
predicted by a Shakura-Sunyaev model, we find that the disk
atmosphere stretches upward, with the photosphere rising to
$\simeq 7H$, in contrast to the $\simeq 3H$ predicted by conventional
analytic models.  This more extended structure, as well as fluctuations
in the height of the photosphere, may lead to departures from Planckian
form in the emergent spectra.  Dissipation is distributed across the region
within $\simeq 3H$ of the midplane, but is very weak at greater
altitudes.  As a result, the temperature deep in the disk interior
is less than that expected when all heat is generated in the
midplane.  With only occasional exceptions, the gas temperature stays very
close to the radiation temperature, even above
the photosphere.  Because fluctuations in the
dissipation are particularly strong away from the midplane, the
emergent radiation flux can track dissipation fluctuations with a
lag that is only 0.1--0.2 times the mean cooling time of the disk.
Long timescale asymmetries in the dissipation distribution can also
cause significant asymmetry in the flux emerging from the top and
bottom surfaces of the disk.
Radiative diffusion dominates Poynting flux in the vertical energy flow
throughout the disk.

\end{abstract}


\section{Context}
    If accretion disks radiate dissipated energy more quickly than 
their material moves inward, simple energy conservation determines 
their effective temperature as a function of radius $r$: it is
\begin{equation}
T_{\rm eff}=\left[\frac{3}{8\pi\sigma}\frac{GM\dot MR(r)}{r^3}\right]^{1/4},
\label{eq:teff}
\end{equation}
where $\sigma$ is the Stefan-Boltzmann constant, the central mass is $M$, the
accretion rate is $\dot M$, and $R(r)$ is a correction factor (close to unity
at large radius) that accounts for
the effect of the net angular momentum flux through the disk and relativistic
corrections.  Inside the disk, the temperature is often supposed to increase
$\propto \tau^{1/4}$ (for optical depth from the surface $\tau$), in keeping
with conventional local thermodynamic equilibrium (LTE) 
stellar atmosphere theory (e.g., as in Shakura \& Sunyaev 1973).
This guess about the internal temperature gradient is founded, however,
upon the tacit assumption that the outgoing flux is created mostly near
the midplane, which may or may not be true in real disks.  Unfortunately,
despite all the effort that has been expended upon accretion disk dynamics,
there has been as yet only very limited attention given to their
thermal structure \citep{Brandenburg et al.(1995), Turner(2004)}

      The true temperature profile matters in a number of ways.  It is, of
course, the fundamental parameter governing vertical hydrostatic balance,
whether the pressure is dominated by gas or radiation.  Thus, the disk's
thickness depends upon it.  In addition, for known effective temperature
(see equation 1), the spectrum emergent from any particular ring depends most
strongly on the gravity at the photosphere.  In thin disks, the vertical 
component
of gravity increases linearly with height, so the temperature profile governs
this quantity as well.  Whether conventional plane-parallel, time-steady
stellar atmosphere theory even applies to accretion disks also depends on
whether the temperature (and density) structure is sufficiently homogeneous
and stationary, one more open question.

     Still another aspect of accretion disks for which knowledge of the 
vertical temperature profile is vital is the origin of hard X-ray emission.  
A very-nearly
ubiquitous property of accretion disks around black holes, it appears on
phenomenological grounds to find its origin in a ``coronal" region somewhere
near the disk surface.  It is widely assumed that somehow the vertical
temperature profile must
accommodate a sharp upward rise above the photosphere, as the highest
$T_{\rm eff}$ even in disks around relatively low-mass black holes
($\sim 10 M_{\odot}$ is merely $\sim 10^7$~K, whereas the observed hard
X-rays demand $T_e \sim 10^9$~K.

    With the recognition that accretion is due to magnetohydrodynamic 
(MHD) turbulence driven 
by an underlying magneto-rotational instability 
\citep[MRI: see, e.g., the review by][]{Balbus & Hawley(1998)}, 
the path toward answering these questions has
been laid out, at least in principle:  Track the dissipation that must
occur at the short lengthscale end of the turbulent cascade; use known
radiation physics to relate heat injection to photon production; compute
photon transfer to the surface.  The result of following this program will
be a complete description of the disk equation of state and thermodynamics.

    It is the object of this paper to report a step toward carrying
out this project.  As we will describe at greater length in the body of the
paper, we have simulated a shearing-box section of a gas-dominated
accretion disk in a manner that captures all numerically-dissipated
energy.  We suppose that this energy is transformed into local
heating, and then solve the radiation transfer problem in the flux-limited 
diffusion (FLD) approximation.  
With this method, we are able to find the internal
profiles of dissipation rate and temperature, as well as gas density
and other interesting physical quantities.  We are also able to study
the character of fluctuations, both spatial and temporal.

    Of all the many previous numerical studies of accretion disk 
properties that
have been done to explore the consequences of MRI-driven MHD turbulence,
the work by \citet{Miller & Stone(2000)} is the one that most closely
approached these questions before the present effort.  In that paper, they
explored the vertical structure of MHD turbulence within a shearing-box,
but did so making a simplifying assumption about the gas thermodynamics: that
the equation of state was isothermal.  We will use their results as
a standard of comparison in order to clarify which of our findings
depend on explicit calculation of the gas's thermodynamics, and which
may be found on the basis of simpler arguments.

\section{Calculation}
The subject of our simulation is a radially thin slice of a 
gas-dominated accretion disk.  We approximate the dynamics
of this slice by a shearing box.  That is, if we denote the radial
coordinate by $x$, the azimuthal coordinate by $y$, and the vertical
coordinate by $z$, we assume there are underlying
(unsimulated) dynamics that produce an azimuthal velocity 
$v_y = -(3/2)\Omega x$,
where $\Omega$ is the orbital frequency and the zero-point of $x$ is
in the center of the box.  Because
this shearing-box is meant to approximate a section of an orbiting
disk, we also assume a Coriolis force and appropriate gravitational
tidal forces.

\subsection{Basic Equations and their Solution}\label{sec:basiceq}
The equations we solve to describe the dynamics of this box
are the frequency-averaged equations 
of radiation MHD in the FLD approximation:
\begin{eqnarray}
  &&\frac{\partial\rho}{\partial t}+\nabla\cdot\left(\rho\bm{v}\right) = 0\\
  &&\frac{\partial \rho\bm{v}}{\partial t}
  +\nabla\cdot\left(\rho\bm{v}\bm{v}\right)
  = -\nabla(p+q) + \bm{j}\times\bm{B}
  +\frac{\chi\rho}{c}\bm{F} \nonumber\\
  &&\qquad - 2\rho\left(\Omega\hat{\bm{z}}\right)\times\bm{v}
  + 3\rho\Omega^2x\hat{\bm{x}}-\rho\Omega^2z\hat{\bm{z}}\label{eq:mom}\\
  &&\frac{\partial E}{\partial t}+\nabla\cdot\left(E\bm{v}\right)
  = (4\pi B-cE)\kappa\rho-\nabla\bm{v}:\mathsf{P}
  -\nabla\cdot\bm{F}\label{eq:erad}\\
  &&\frac{\partial e}{\partial t}+\nabla\cdot\left(e\bm{v}\right)
  = -(4\pi B-cE)\kappa\rho-\left(\nabla\cdot\bm{v}\right)(p+q)
       + Q
\label{eq:eint}\\
  &&\frac{\partial \bm{B}}{\partial t}+\nabla\times\bm{E}=0\label{eq:indc}\\
  &&\bm{F}=-\frac{c\Lambda}{\chi\rho}\nabla E\label{eq:flux}
\end{eqnarray}
The quantities $\rho$, $e$, and $\bm{v}$ are the density, internal energy, 
and velocity field of the gas. The pressure of the gas $p$ 
is related to the internal energy by $p=(\gamma-1)e$ 
with the adiabatic index $\gamma$ (here $\gamma=5/3$ is assumed).
$q$ is the stress associated with a small artificial bulk viscosity
employed, in the usual manner, to suppress post-shock ringing by
capturing the shock's entropy generation.
$Q$ represents the rate at which the internal energy
must be changed in order for total energy to be conserved 
(see Appendix \ref{app:cnsv}).
$\bm{B}$ is the magnetic field, 
$\bm{j}=\nabla\times\bm{B}/(4\pi)$ is the current density, and 
$\bm{E}=-\bm{v}\times\bm{B}$ is the electric field.

The radiation field is described by 
the energy density $E$, flux $\bm{F}$, and pressure tensor $\mathsf{P}$. 
The equations are closed with the relation $\mathsf{P}=\mathsf{f}E$, where 
$\mathsf{f}$ is the Eddington tensor, which is defined in terms of the
scalar Eddington factor $f$:
\begin{equation}
\mathsf{f} = (1/2)(1-f)\mathsf{I} + (1/2)(3f-1)\bm{\hat n}\bm{\hat n}.
\end{equation}
The Eddington factor is in turn assumed to depend on the flux limiter
$\lambda$ and the dimensionless opacity parameter $R\equiv|\nabla E|/(\chi \rho E)$
through the relation,
\begin{equation}
f = \lambda + \lambda^2 R^2
\end{equation}
\citep{Turner & Stone(2001)}, where $\lambda = (2+R)/(6+3R+R^2)$ 
\citep{Levermore & Pomraning(1981)}.
Assuming LTE, the radiation emission rate is simply the 
Planck function, $B$.  Appropriately averaging the frequency-dependent
free-free opacity, the Planck-mean opacity $\kappa$ depends on the
other parameters as $3.7\times10^{53}\sqrt{\rho^{9}/e^7}$ cm$^2$ g$^{-1}$
(one of the reasons we choose a comparatively small central mass,
$6.62M_{\odot}$, for this
simulation is to reduce the contribution of other mechanisms to the
absorptive opacity).  Both free-free absorption and Thomson scattering
are included in the Rosseland-mean opacity $\chi$
($=1.0\times10^{52}\sqrt{\rho^{9}/e^7}+0.33$ cm$^2$ g$^{-1}$). 

   For the problem studied in this paper, the thermodynamics and
vertical structure of a gas pressure-dominated disk, the radiation
force term in equation~\ref{eq:mom} might seem superfluous.  We retain
it for two reasons.  First, although in the conditions treated here
the radiation force is relatively small, it still contributes at the
tens of percent level.  Second, one of our long-term goals in this
effort is to study disks in which radiation forces are more important,
so we wish to develop techniques capable of describing them.

    It is also worth commenting that there are some problems in which
the gas's radiative losses could be treated more simply.  If, for
example, the ion temperature is elevated far above the electron temperature
\citep{Ichimaru(1977), Rees et al.(1982), Narayan & Yi(1994)}, the
density in the disk can be so low that the optically thin limit can
apply.


   The tool we use to solve these equations is the latest version of the
radiation/MHD code first described in \citet{Stone & Norman(1992a), 
Stone & Norman(1992b)}, and then successively modified by
\citet{Turner & Stone(2001)}, \citet{Turner et al.(2003)} 
and again by \citet{Turner(2004)}.
We have further improved it.  The most important change in terms of
its applicability to this scientific problem is that it now conserves
energy quite accurately by means of a special internal energy update
scheme designed to ensure that all numerical losses of kinetic and
magnetic energy are captured in the form of heat added to the gas
internal energy (details in Appendix \ref{app:cnsv}).  In this way,
these equations mimic the much more complex physics of magnetic
reconnection and other kinetic processes.

In addition, we have also made several technical improvements.
Two are worth noting here.  First, instead of the Alternating Direction
Implicit (ADI) method used by \citet{Turner & Stone(2001)} to solve the
radiation diffusion equation, we now employ the Gauss-Seidel
method accelerated 
by a full multi-grid method.  When the ADI method was used 
with a shearing box boundary condition, maintaining stability restricted
the time step $\Delta t$ to be shorter than the diffusion 
time step.  Our code is free from this restriction 
because the Gauss-Seidel method solves the diffusion equation in a 
directionally unsplit way.  The Courant number for the radiation 
diffusion can be as large as $\sim 10^2$ in our simulation.

Second, in every grid-based hydrodynamics simulation with explicit
time-advance, the Courant condition puts an upper limit on the
time-step that can be used.  In this case, we evaluate
this limit in terms of the time for an acoustic wave supported
by gas, magnetic, and radiation pressure to cross a grid-zone.
This time is much shorter in the low-density regions above the
photosphere than in the midplane because the ratio of radiation
pressure to density is very large there.  Consequently, the
time-step is governed primarily by the lowest-density zones.
If the density there were permitted to be as small as the dynamics
demanded, a prohibitive number of time-steps would be required
in order to complete the simulation.  In the closely-related
work of \citet{Turner(2004)}, a lower bound of $2 \times 10^{-3}$ of the
maximum density in the initial state was placed on the density in
order to prevent such a computational slow-down.  However,
invocation of the density floor effectively means injection of
energy into the problem because the newly-created mass is
given the local velocity and implicitly acquires the local potential energy.
To enhance the quality of our energy
conservation while still limiting the total computing
time required to a reasonable amount, we set this floor
at $10^{-5}$ of the midplane density.  At this value, the total
amount of energy injected due to the density floor
is $\simeq 0.9\%$ of the total heat dissipation over the 60
orbits of the simulation.  Further details
about this procedure (and the smaller effects of some other variable
caps) may be found in Appendix~\ref{app:arti}.


\subsection{Parameters}
   The physical conditions in such a slice are determined by three
parameters: the central black hole mass $M$, the central radius of the
slice $r$, and the surface density $\Sigma$.  The values we
have chosen for them are meant to be representative of conditions in an
accretion disk around a Galactic black-hole binary relatively far from
the black hole itself: $M= 6.62M_{\odot}$, $r = 300r_g$, and the
surface density predicted by an $\alpha=0.03$ Shakura-Sunyaev disk model
with an accretion rate $\dot M$ that would yield a total luminosity of
$0.1L_E$ if the radiative efficiency in rest-mass units were 0.1.
The effective temperature at the surface of such a disk segment is
$5.3 \times 10^5$~K.  These parameters can be combined to define a
characteristic scale-height $H$ after estimating the temperature
rise toward the center of the disk.
In such a model, the relation between surface density and accretion rate is
\begin{equation}
\Sigma_0 = \left(\frac{\pi \sigma}{3\chi}\right)^{1/5}
           \left(\frac{\mu}{\pi \alpha k}\right)^{4/5}
           \dot M^{3/5} \Omega^{2/5} R^{3/5}(r).
\label{eq:sigma}
\end{equation}
The symbols $\mu$, $k$, and $\sigma$ in this expression are the
mean mass per particle, the Boltzmann constant, and the Stefan-Boltzmann
constant, respectively.
Numerical values for these and a number of other parameters of interest
are collected in Table \ref{tab:param} (note that in the shearing-box
approximation, $R(r) \equiv 1$).

\begin{deluxetable}{cccl}
\tablecolumns{4}
\tablecaption{Simulation Parameters\label{tab:param}}
\tablehead{
\colhead{Parameter} & \colhead{Definition} & \colhead{Value} & \colhead{Comment}
}
\startdata
$M$          &$6.62M_\odot$              &$1.31\times10^{34}$~g          & mass of central black hole\\
$\dot{M}$    &$0.1(L_E/c^2)/\eta$        &$7.52\times10^{17}$~g~s$^{-1}$         & estimated mass accretion rate       \\
$\alpha$     &                           &$0.03$              & estimated stress/pressure ratio     \\
$r$          &$300(GM/c^2)$              &$2.93\times10^{8}$~cm          & radius                    \\
$\Omega$     &$\sqrt{GM/r^3}$            &$5.90$~s$^{-1}$        & orbital frequency         \\
$T_{\rm orb}$&$2\pi/\Omega$              &$1.06$~s         & orbital period            \\
$H$          &$c_{\rm g}/\Omega$\tablenotemark{a} 
                                 &$3.53\times10^6$~cm        & disk scale-height         \\
$F_0$        &$(3\dot{M}\Omega^2)/8\pi$  &$3.12\times10^{18}$~erg~cm$^{-2}$~s$^{-1}$& estimated radiation flux at surface \\
$\Sigma_0$   &eq.[\ref{eq:sigma}]        &$9.89\times10^4$~g~cm$^{-2}$    & surface density           \\
$\rho_0$     &$\Sigma_0/(2H)$            &$1.40\times10^{-2}$~g~cm$^{-3}$ & mean density \\
\enddata
\tablenotetext{a}{$c_{\rm g}$ is the gas sound speed at the effective temperature.}
\end{deluxetable}

\subsection{Initial Conditions}
   Our initial condition is (almost) the Shakura-Sunyaev equilibrium
corresponding to our chosen parameters.  In this equilibrium,
the gas and the radiation are assumed to be in a planar 
hydrostatic and thermal equilibrium state everywhere below the
photosphere.  Our initial condition differs from the Shakura-Sunyaev
assumptions in taking the local dissipation rate proportional to 
the logarithm of the column density to the surface, a dissipation profile
adopted on the basis of some previous numerical experiments. 
Under these assumptions, the basic equations (\ref{eq:mom}), 
(\ref{eq:erad})+(\ref{eq:eint}), and (\ref{eq:flux}) 
are reduced to 
\begin{eqnarray}
&&-\frac{dp}{dz}+\frac{\chi\rho}{c}F-\rho\Omega^2z=0 \\
&&\frac{dF}{dz}=-\frac{F_0}{\ln\tau_0}\frac{d\ln\Sigma}{dz}\\
&&F=-\frac{c}{3\chi\rho}\frac{dE}{dz},
\end{eqnarray}
where $\tau_0\equiv\int_{0}^{z_{\rm surface}}\chi\rho dz=\chi\Sigma_0/2$ 
is the total optical depth
and we assume the opacity is entirely Thomson scattering. 
The boundary condition on this set of equations is that the radiative
flux at the photosphere ($\tau=1$) of the disk
must match the total dissipation rate (eq.[\ref{eq:teff}]), or $F_0$.
Lastly, the gas is also assumed to be in thermal equilibrium with the 
radiation ($E=aT^4$). 

Outside the photosphere, the flux is constant ($F=F_0$), 
the gas density is set to the density floor (see \S~\ref{sec:basiceq}), 
and its temperature is constant at the effective temperature, $4.8 \times 10^5$~K.
Because its total pressure is then very nearly constant with height, 
this gas has almost no support against gravity.

\begin{figure}
\epsscale{0.75}
\plotone{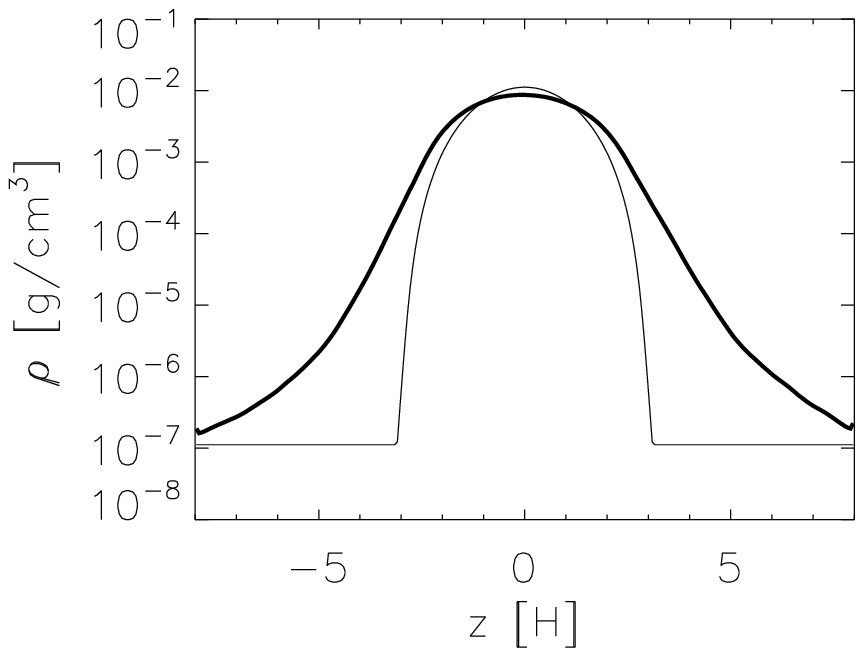}
\caption{Horizontally-averaged density: initial condition (thin curve) and  
time-averaged profile (thick curve).
\label{fig:densprofile}}
\end{figure}

\begin{figure}
\epsscale{1.2}
\plottwo{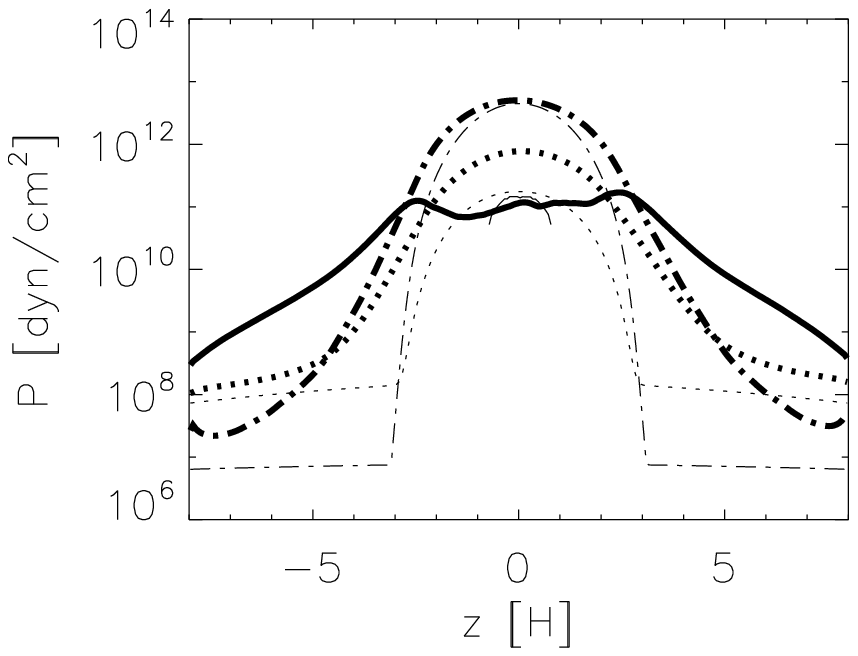}{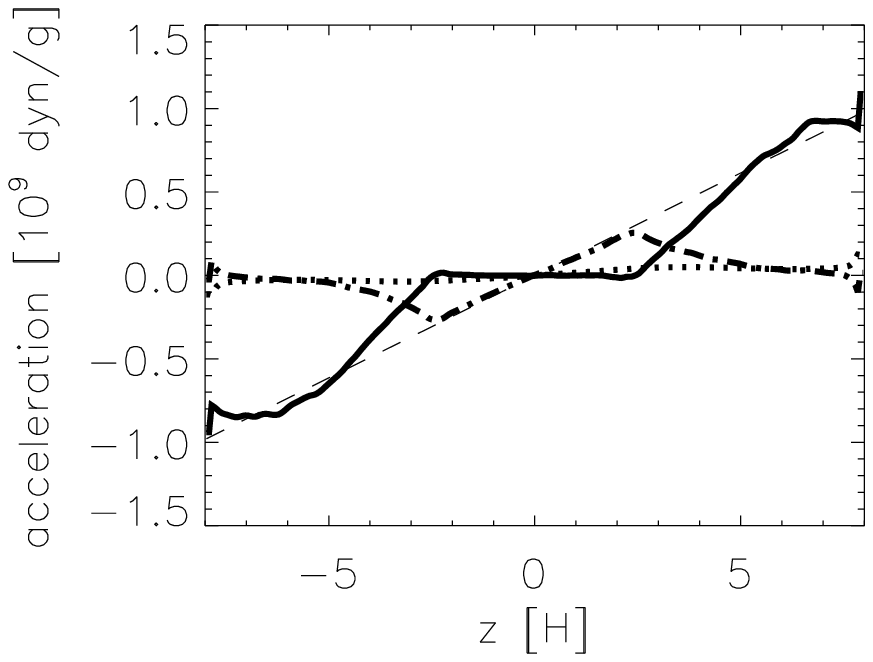}
\caption{Left panel: horizontally-averaged pressure.  All initial condition
curves are thin, all time-averaged 
curves are thick.  In each case, gas pressure  is represented by a
dot-dash curve, magnetic pressure  by a solid curve, and
radiation pressure by a dotted curve.  Right panel: time- and
horizontally-averaged contributions to net vertical acceleration. Gravity
is shown by a dashed curve, other quantities by the same line-styles as
in the left-hand panel. 
\label{fig:pressprofile}}
\end{figure}

The shapes of the initial gas density and 
pressure profiles are displayed in Figures~\ref{fig:densprofile} and
\ref{fig:pressprofile}.
As expected, the ratio of radiation pressure to gas pressure in the
midplane is small, $0.039$ in the initial state, placing this equilibrium
well within the gas-dominated regime.  With such a large surface density, 
the disk is also very optically thick: its (half)-Thomson depth is 
$\tau_0=1.63\times10^4$.  Although the absorptive opacity is never more than
$\simeq 6\%$ of the Rosseland mean, the effective optical depth
to the midplane is also so large
($\int_{0}^{z_{\rm surface}}\sqrt{\kappa\chi}\rho dz=4.04\times10^3$)
that we expect the radiation to be well-coupled thermally to the gas 
except on very short lengthscales and at very high altitude.  This
expectation is strongly vindicated in the simulation 
(Fig.~\ref{fig:temperature}).

The initial configuration of the magnetic field is a twisted azimuthal 
flux tube of circular cross section with radius $0.75H$ whose center is
placed at $x=z=0$. The field strength in the tube is uniform at
$2.16\times10^6$~G, and its maximum poloidal part is $5.40\times10^5$~G,
corresponding to an initial plasma $\beta\equiv p/(\bm{B}^2/8\pi) = 24$. 
The corresponding maximum MRI wavelength in the midplane
is $\lambda_{\rm max}\equiv(8\pi/\sqrt{15})(v_A/\Omega)=1.59\times10^6$ cm, 
which is resolved with $7.2$ grid zones 
\citep{Balbus & Hawley(1998)}.

The calculation is begun with a small random perturbation in the poloidal 
velocity. The maximum amplitude of each velocity component is $1$\%
of the local magnetosonic speed $c_s$, in which we include radiation
in the total pressure.

\subsection{Grid and Boundary Conditions}\label{sec:bndc}
The computational domain extends $2H$ 
along the radial direction, $8H$ along the orbit, and $8H$ on either 
side of the midplane. It is divided into $32\times 64\times 256$ cells
($x \times y \times z$) with constant grid size
$\Delta x=\Delta z= H/16 = \Delta y/2$.
The grid is staggered, with scalar quantities
($e$, $E$, and $\rho$) defined at zone-centers and vector quantities 
($\bm{v}$, $\bm{B}$, and $\bm{F}$) defined on zone-surfaces.

The azimuthal boundaries are periodic and the radial boundaries are 
shearing-periodic in the local shearing-box approximation 
\citep{Hawley Gammie & Balbus(1995)}.  We desire the vertical
boundaries to be outflow (free) boundaries; that is, we endeavor
to determine the values in the ghost cells 
so that mass, momentum and energy fluxes across 
the boundaries are the same as those across the adjacent cell surfaces.
However, the vertical disk structure is approximately hydrostatic,
so motions across the top and bottom boundaries are in general
subsonic.  Imperfections in the boundary conditions can therefore
act as noise sources for perturbations able to travel into the problem
area.  We found in practise that many natural
realizations of outflow boundary conditions for this problem suppressed
noise-generation sufficiently well that little was
seen for tens of orbits, but dramatically elevated values of
one or more of the code variables would then suddenly appear
in a small number of cells and disrupt the simulation.  The particular
version of outflow boundary conditions we present in the
next few paragraphs was able to suppress such artifacts for
the entire 60 orbit span of this simulation.

For solving advection equations (steps 4 to 7 in Appendix
\ref{app:cnsv}), the usual approach---a constant (zero slope)
projection of variables into the ghost cells---works well.
For solving the radiation diffusion equation (step 3 in
Appendix \ref{app:cnsv}), it is better to set $E$ in the 
ghost cells so that the diffusion flux
$F_z=-c\Lambda/(\chi\rho)dE/dz$ is constant
(details are described in Appendix \ref{app:ebndc}).
We tried both time-explicit and time-implicit versions of
this method and found that (somewhat surprisingly) the
explicit version was less likely to cause problems.

To prevent energy inflows, we adopt two additional conditions:
(a) the $z$-component of the velocity at the boundary 
surface is set to zero when it is negative (positive) at 
the top (bottom) boundary (see Appendix \ref{app:arti}). 
(b) the radiation energy in the ghost 
cell is set equal to that in the adjacent cell when the former 
becomes larger than the latter.

For the magnetic field, we build upon the usual scheme in
which the electric field is extrapolated into the ghost cells
and the induction equation (eqn.~\ref{eq:indc}) is solved
in those cells to find the magnetic field.  This approach
ensures that $\nabla\cdot\bm{B}=0$ even
in the ghost cells (the CT algorithm ensures it is zero in the
entire problem area).  It does not, however, 
guarantee that the magnetic field varies smoothly across the
boundaries.  To forestall jumps in the magnetic field
intensity across the boundary, we introduce a
small but finite resistivity into the ghost cells.  This
device also prevents the development of field-strength
discontinuities at the boundary large enough to cause
code crashes.  After some experimentation, we found that
these goals could be achieved with a ghost-cell resistivity set to 
$0.13\times{\min({\Delta x}^2,{\Delta y}^2,{\Delta z}^2)}/\Delta t$.  
Note that our electric field extrapolation
does not restrict the sign of the Poynting flux across the boundary
(see section \ref{sec:ecnsv}).

\section{Results}
    The simulation ran for 60 orbits.  Transients persisted
for 5--10 orbits, and thereafter most properties maintained
a statistically steady state.  For example, during the
period from 10 to 60 orbits, the outgoing flux varied within
a total range of a factor of two, with an rms fractional variation
of only 14\%.  We therefore regard the structure as having achieved a
well-defined steady-state, and we begin by describing its
time-average properties. In the following, the time average is done for 
50 orbits, from $t=10$ orbits to $t=60$ orbits.

\subsection{Mean Vertical Profiles}
    Our initial condition is (almost) a conventional model for a gas
pressure-dominated disk: a (nearly-) Gaussian profile of gas density with
characteristic scale $H$.  However, the time-averaged density profile is
considerably shallower in slope (Fig.~\ref{fig:densprofile}).  From a
height $|z| \simeq 2H$ out to $\simeq 8H$, it is better described by an
exponential with a slowly changing length-scale: the density
scale-height is $\simeq 0.3H$ for $H \lesssim z \lesssim 5H$ and
stretches to $\simeq 0.4H$ at greater altitude.  This result echoes
what was found by Miller \& Stone (2000), but is different
in detail: we find a broader region of slowly-declining density near
the midplane (presumably do to our higher temperature there)
and slightly steeper exponential decline outside that
region.\footnote[1]{Note that Miller \& Stone's definition of the
scale-height $H$ is $\sqrt{2}$ times larger than ours, so their
vertical extent of $\pm 5$ scale-heights corresponds to $\pm 7H$
in our units.}

Because the density falls relatively gradually
with height, the mean altitude of the
photosphere (defined as the point where the Eddington factor $f = 0.5$:
see \S~\ref{sect:cooling} for a more detailed discussion)
is almost at the top of the box, $|z| \simeq 7.3H$.  Thus,
although most of the mass is contained within $\pm 2H$ of the midplane,
in other respects the disk should be thought of as considerably thicker.

    Clearly, something other than thermal gas pressure must support 
this extended atmosphere.  What that is is revealed in 
Figure~\ref{fig:pressprofile}, where we
display the profiles of gas pressure ($p=2e/3$), magnetic 
pressure ($p_{\rm mag}\equiv\bm{B}^2/8\pi$), 
and radiation pressure ($p_{\rm rad}\equiv E/3$), as well as the
contribution of each of them to support against gravity.  Although both
the radiation pressure and the magnetic pressure are small compared
to the gas pressure throughout the initial condition, both become
greater than the gas pressure above $\simeq 3H$ in the equilibrium,
with the magnetic pressure generally a few times greater than the
radiation pressure.  Deep inside the
disk, hydrostatic support is mostly due to gas pressure.  Although 
at higher altitudes the radiation pressure is comparable to the magnetic
pressure, and both are larger than the gas pressure, the dominant vertical
force is magnetic.  The reason is that near the photosphere, where the
radiation pressure is relatively large, its gradient is small.  
Thus, we find that the disk possesses an extended sub-photospheric 
magnetically-dominated atmosphere whose thickness is more than twice 
that of the gas pressure-supported disk body.

    Although the magnetic field dominates at higher altitudes, the radiation
pressure grows to exceed the magnetic field energy density near the midplane.  
From its initial midplane
ratio of $\simeq 0.04$, $p_{\rm rad}/p$ grows to $\simeq 0.2$ in the steady
state, about a factor of 6 greater than $p_{\rm mag}/p$ at that location.
Similarly, the plasma $\beta$ (which was 24 in the midplane initially)
grows to $\simeq 40$ there, but declines steeply with height, reaching a low
of about 0.03 at $|z| \simeq 7H$. If the pressure used in the plasma $\beta$
is redefined as the sum of radiation plus gas pressure, the
minimum $\beta$ is $\simeq 0.15$.  These values are very similar to
those previously found by \citet{Miller & Stone(2000)}, a midplane
$\beta \simeq 40$ and $\beta(7H) \simeq 0.01$.

\begin{figure}
\epsscale{0.75}
\plotone{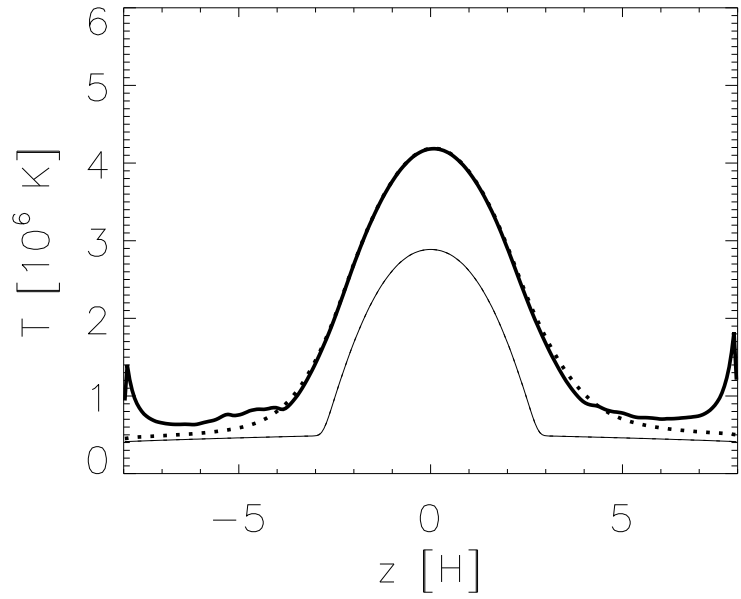}
\caption{Temperature as a function of height.
Gas temperature is shown by the solid curves, heavy for the time-averaged
profile, thin for the initial condition; time-averaged
radiation temperature is shown by the dotted curve.
\label{fig:temperature}}
\end{figure}

    The gas temperature declines from a peak of about $4.5 \times 10^6$~K at
the midplane to a minimum $\simeq 7 \times 10^5$~K, achieved in the range
$5H \lesssim |z| \lesssim 7H$ (Fig.~\ref{fig:temperature}).  Within
$\simeq 5H$ of the midplane, LTE is an excellent approximation, as
the gas and radiation temperatures are very close.  In the outer
most of that zone, the $\tau^{1/4} T_{\rm eff}$ approximation is fairly
close to the actual temperature, but it overestimates the temperature
by $\simeq 20\%$ within $\simeq H$ of the midplane.  This overestimate
is due to the finite vertical spread in the dissipation distribution
(see section \ref{sect:dissipation}).  At large $|z|$, the radiation
temperature approaches the effective temperature (as it must).  The
time-averaged effective temperature is $\simeq 5.3 \times 10^5$~K,
about 10\% higher than the temperature predicted by the Shakura-Sunyaev
model.  However, the gas temperature between $\simeq 5H$ and $7H$ exceeds the
radiation temperature by about 30\%.

   Above $\simeq 7H$,
the gas temperature rises toward our outer boundary.  It is hard
for us to determine how much of that rise is physical, and how much is due to
the guesses required to impose an outer boundary on the magnetic field.
We are led to suspect that much of it is a numerical artifact because
roughness in the magnetic field near the boundary creates a spike in the
time-averaged vertical profile of the dissipation density per unit mass,
likely leading to this abrupt rise in temperature.
However, as Figure~\ref{fig:temperature} also shows, the radiation
temperature decouples from
the gas temperature near the gas temperature minimum, so that the outgoing
radiation flux is hardly affected by the gas temperature spike near the
boundary.

\begin{figure}
\epsscale{1.0}
\plotone{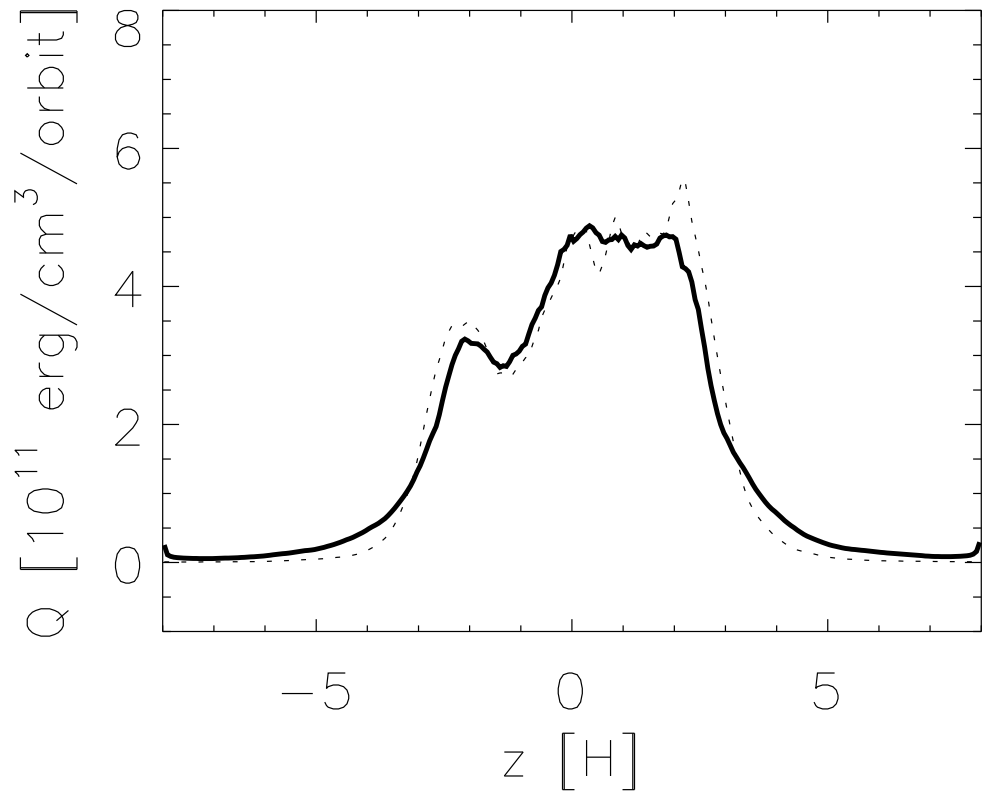}
\caption{Time- and horizontally-averaged
dissipation as a function of height within the disk (solid curve).
Stress multiplied by $(3/2)\Omega$ is shown with a dotted curve.
\label{fig:dissipation}}
\end{figure}

    Lastly, we turn to the time-averaged stress.  Even after both horizontal-
and time-averaging, the stress profile still shows small-scale fluctuations
and a distinct asymmetry---the peak is at $z\simeq 2H$, and is about 50\%
higher than the level at $z\simeq -2H$ (Fig.~\ref{fig:dissipation}).
In fact, we find that at any
given time the magnetic field intensity and stress tend to be 
asymmetric about the midplane, with both a few times greater in one half
than in the other.  These asymmetries last for periods $\sim 10$~orbits before
flipping to the other direction.

     Like many other quantities, there is also a sharp contrast
between the stress in the disk body region ($|z| \lesssim 3H$) and in
the corona.  Within the disk body region, the stress distribution is
roughly independent of $z$, but outside that region, it declines
sharply.  Virtually all the stress is created in the disk body,
with only about 10\% above $|z| = 3H$ and just 1\% more than $5H$
from the plane.  This pattern is qualitatively similar to the
results of \citet{Miller & Stone(2000)}, who also found an
approximately flat-topped stress distribution that extended a
few scale-heights from the midplane; the only contrast is that
in their work the drop toward the outside was not as steep.

     Averaged over the length of the simulation, the total stress in the
problem area was large enough to drive an accretion rate of
$1.1 \times 10^{18}$~g~s$^{-1}$, 44\% greater than our initially
estimated accretion rate.  This is why $T_{\rm eff}$ is 10\% higher
than as predicted by the Shakura-Sunyaev model, of course.
Phrased in terms of the traditional $\alpha$
measure (ratio of vertically-integrated stress to vertically-integrated
total pressure), we find $\alpha \simeq 0.02$ (note that the time-averaged
vertically-integrated pressure is greater than in the initial
condition).  In evaluating this finding, it should be borne in mind
that global disk simulations \citet{Hawley & Krolik(2001)} generally
find rather larger mean $\alpha$ measures than shearing-box
simulations like this one.  In addition, we find that this stress
is very non-uniformly distributed.  Measuring it in pressure units,
it rises from $\simeq 0.01$ in the midplane to $\simeq 1$ at $|z| > 6H$.


\subsection{Thermal balance}
\subsubsection{Net energy flow}\label{sec:ecnsv}

\begin{figure}
\epsscale{0.75}
\plotone{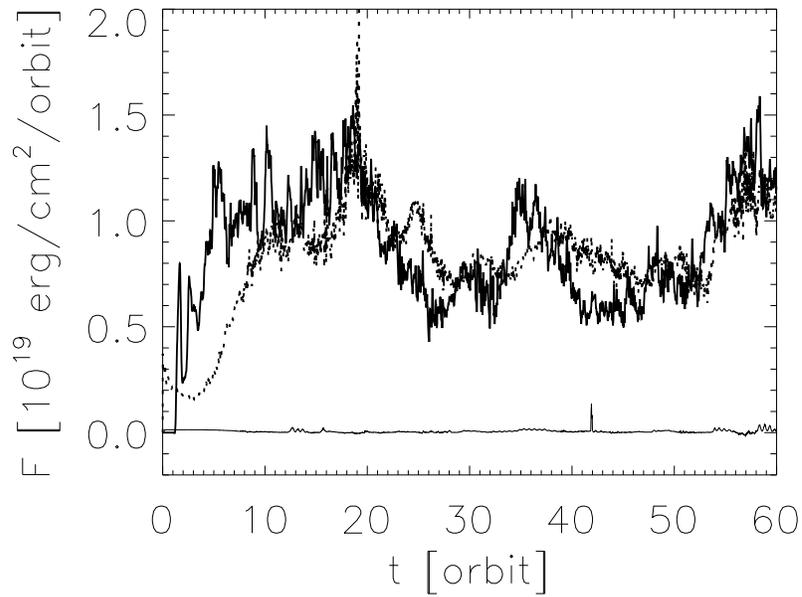}
\caption{
Work done on radial surfaces (solid), 
outgoing energy through top and bottom boundaries (dotted), 
and artificial energy injection, i.e., $S_1 + S_2 + S_3$ as defined in
Appendix~\ref{app:arti} (thin solid), 
all as functions of time.
\label{fig:energycons}}
\end{figure}

     The history of energy flows into and out of the shearing box is
displayed in Figure~\ref{fig:energycons}.  The only physical source of
energy for the problem volume is the work done on the inner and outer radial 
surfaces by magnetic and Reynolds stresses (about 77\% magnetic), but
there is also a small
non-physical injection of energy whenever the density floor, 
energy floor, and/or velocity cap is invoked (see Appendix \ref{app:arti}).  
Energy loss in the simulation is entirely physical, and may occur
through any of three channels: radiation, matter, and Poynting flux.
Of these, radiation diffusion flux completely dominates the
others: the time-averaged fraction in each flux at the outer boundaries
is 99.71\% (radiation--diffusion), 0.14\% (radiation--advection), 0.25\% (matter),
-0.10\% (Poynting flux: it can be negative because we do not prohibit
it in the boundary conditions; see section \ref{sec:bndc}.)  Given
the complete dominance of radiation as the energy output channel, it
should not be surprising that its history tracks very closely the
history of work done on the box.  The flux cannot precisely follow
the work, however, because it takes a finite time ($\sim 1$ orbit)
for work to be transformed into heat and because it takes more time
(in this case up to $\sim 10$ orbits) for the photons to diffuse out.  The
``lightcurve" is therefore a delayed and smoothed version of the work
(see the following subsection).

\begin{figure}
\epsscale{0.8}
\psfig{file=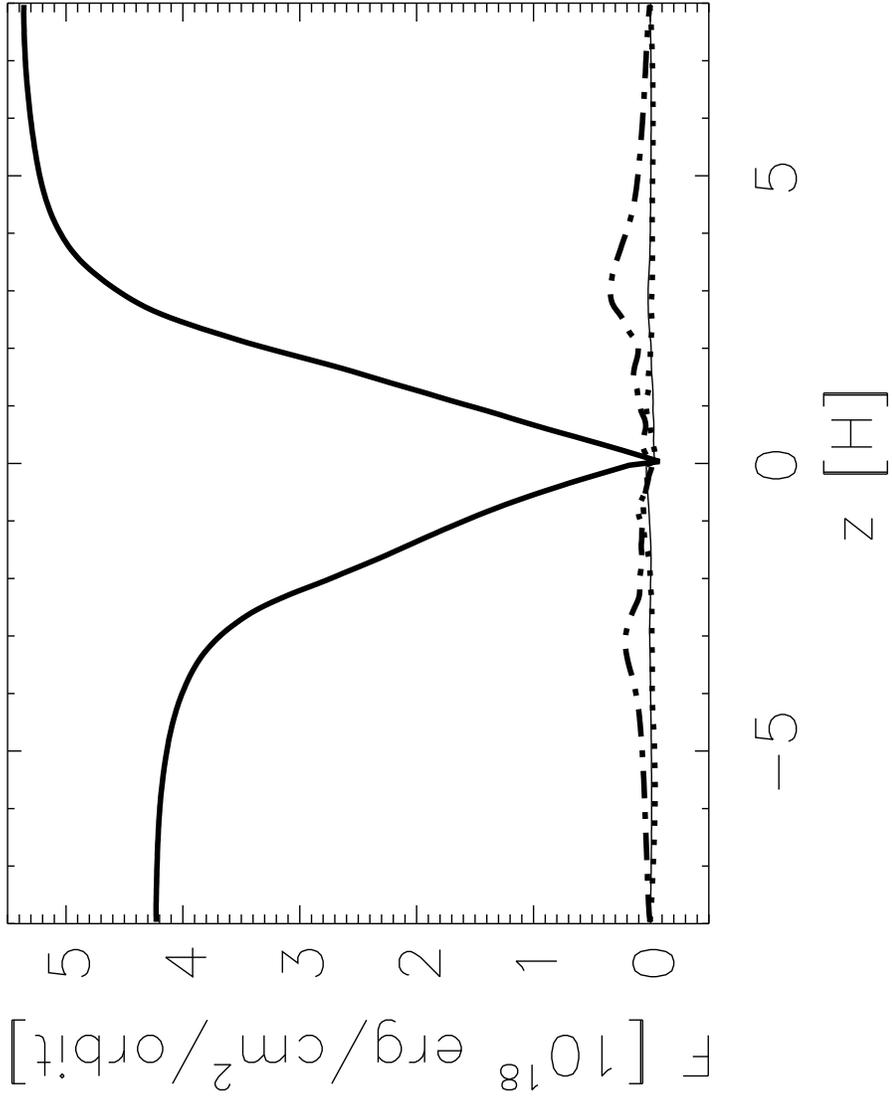,angle=90,width=5.5in}
\caption{The outward energy flux (i.e., flux directed away from the midplane)
carried by radiation (solid curve), Poynting flux (dash-dot curve), and
convective gas motions (dashed curve).  The units of flux are
erg~cm$^{-2}$~orbit$^{-1}$.
\label{fig:fluxprofiles}}
\end{figure}

   Although the vertical radiation flux is always much larger than either the
Poynting flux or the energy flux carried by gas motions,
the ratios between the fluxes of these three channels change systematically
with height (Fig.~\ref{fig:fluxprofiles}).  In the inner half of the
disk (i.e., $|z| \lesssim 3H$--$5H$), Poynting flux is present at
an interesting but low level---$\simeq 5$--8\% of the radiation flux; at
greater height, the Poynting flux becomes much weaker.  By the time the edge
is reached, the Poynting flux is so small that
numerical uncertainties in the boundary conditions lead to its time-average
turning weakly negative, that is, in net it brings energy into the box.
Matter outflow never contributes significantly.  There is also a striking
asymmetry in the magnitude of the flux between the two sides of the disk;
this will be discussed further in \S~\ref{sec:asymm}.

    These results stand in some contrast to the earlier work on stratified
disks by \citet{Miller & Stone(2000)}, where the Poynting flux through the
$|z| = 2H$ surface in their fiducial (toroidal field) simulation was several
times greater, about 25\% of the rate at which stresses did work on
the walls of the box.  In that simulation there were also ``implicit
radiation losses" associated with their isothermal equation of state.

\begin{figure}
\epsscale{0.75}
\plotone{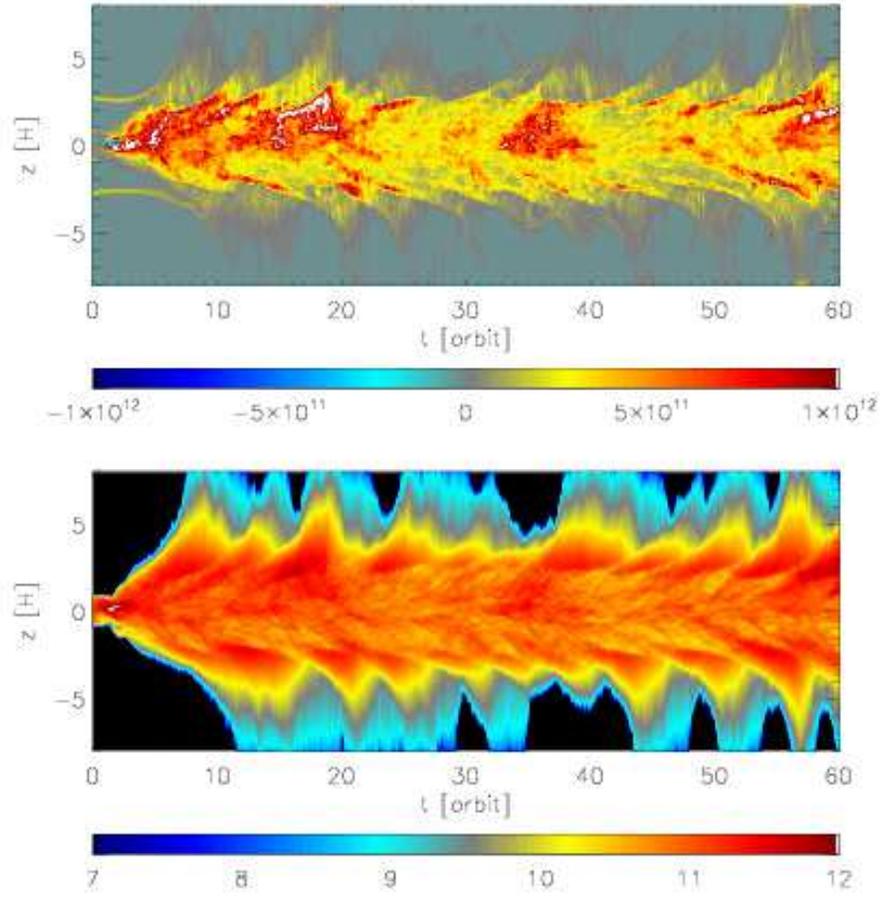}
\caption{Top panel: Horizontally-averaged dissipation rate as a function of
$z$ and $t$.  Bottom panel: Horizontally-averaged magnetic field intensity.
\label{fig:magspacetime}}
\end{figure}

     That the Poynting flux should always be rather smaller than the
total dissipation rate follows from the character of hydrostatic
equilibrium in disks.
The Poynting flux may be thought of as arising from magnetic buoyancy,
and so can be expected to have a magnitude $\sim f c_s |\bm{B}|^2/8\pi$,
where the factor $f$ represents the departure from
hydrostatic balance that permits the upward motion.  On the other hand,
the total dissipation rate is
$\sim (3/2)\phi(z)\Omega s H \langle B_r B_\phi\rangle_0 /4\pi$, where $\phi$
is a profile function that rises from zero in the midplane to unity once
all the dissipation has been accomplished, at $\simeq sH$ (as shown in
the following subsection, $s \simeq 3$ here).  The subscript $0$ labels
the mean stress in the central body of the disk.  Thus, the Poynting flux
fraction is
$$
\sim \frac{\sqrt{2}}{3}{f \over \phi s}{|\bm{B}|^2(z) \over 
                  \langle B_r B_\phi\rangle_0}
$$
because the scale-height $H \equiv c_s/(\sqrt{2}\Omega)$.  In the central disk,
the ratio of the magnetic field intensity to the stress is typically
a few, making the Poynting flux fraction somewhat less than $\sim f$;
at larger altitude, the ratio of the local magnetic field energy density
to the stress in the central disk drops, and the Poynting flux becomes
progressively less important.  Thus, the maximum Poynting contribution
is $\lesssim f$.  In this simulation, we find that the mean rise speed of
magnetic structures (as seen, e.g., in Fig.~\ref{fig:magspacetime})
is $\sim 0.1 c_s$, indicating a maximum Poynting contribution $\sim 10\%$.
It is possible that in other, less settled, circumstances $f$, and
therefore the Poynting flux contribution, might be greater.

\subsubsection{Dissipation rate}\label{sect:dissipation}

\begin{figure}
\epsscale{1.0}
\plotone{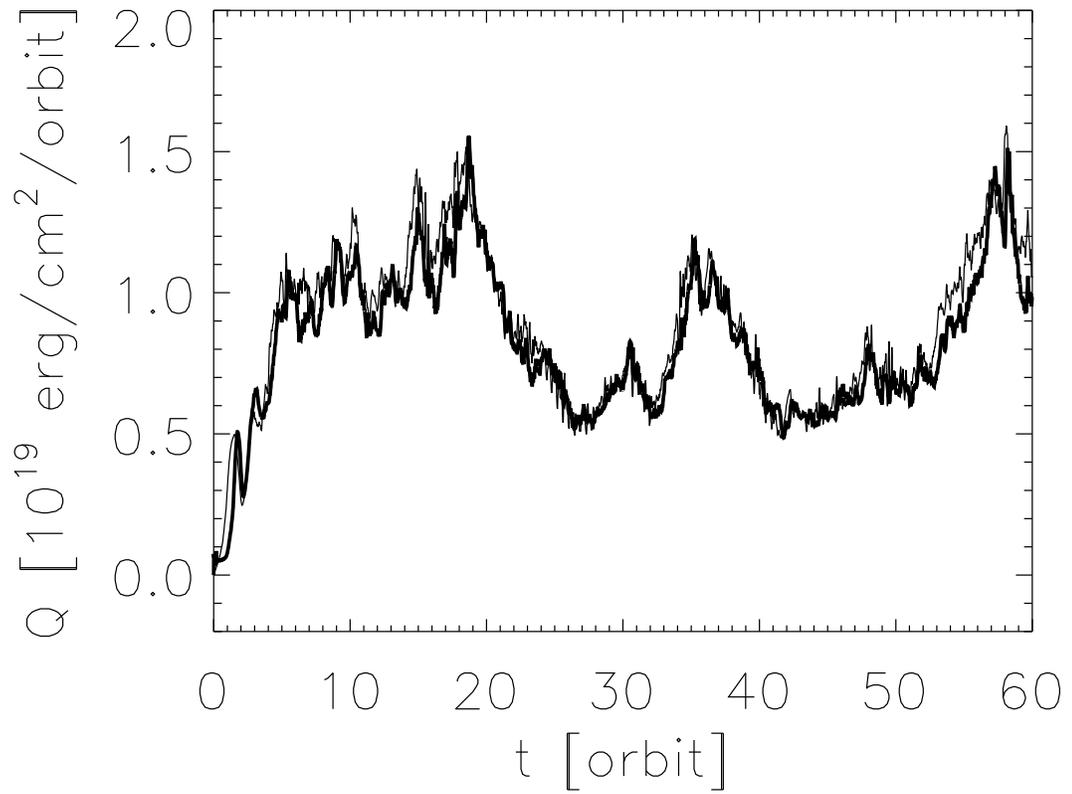}
\caption{Volume-integrated dissipation as a function of time (solid
curve) and volume-integrated $x$--$y$ stress multiplied by $(3/2)\Omega$
(dashed curve).
\label{fig:disshist}}
\end{figure}

    The energy brought into the problem volume by stresses on the radial
surfaces is transformed into magnetic field energy and the kinetic energy of
random fluid motions.  As the turbulent cascade transfers this energy to
smaller and smaller scales, eventually it is dissipated.  A small part of
this dissipation is associated with weak shocks (mostly at high altitude),
but the great majority occurs on the grid-scale and is captured by our
local energy conservation scheme, where it is transferred to gas internal
energy.  Because the step in our energy conservation scheme that captures
magnetic energy dissipation also includes kinetic energy, we cannot fully
separate which contributes how much to the total.  What we can say is
that in rough terms, 70\% of the dissipation comes from the step that mixes
the two, 20\% is clearly associated with the loss of fluid kinetic energy, and
10\% comes from our use of an artificial bulk viscosity.  As we will shortly
demonstrate, however, there is a strong correlation between dissipation rate
and current density; this correlation suggests that much of the mixed-source
70\% is due to magnetic energy.  Two global measures
of this dissipation are shown in
Figures~\ref{fig:dissipation} and \ref{fig:disshist}, its time- and
horizontally-averaged vertical profile and its time-history.

     The total dissipation rate varies by factors of several over time,
but in the mean must result in an effective temperature at the surface 
as given by equation (1).  As can be seen in the figure, its fluctuations
follow very closely upon fluctuations in the total stress.  Cross-correlating
the two curves demonstrates that there is generally a lag of less than an
orbit between stress and dissipation; that is, it takes no more than an
orbit or so for energy injected by shear stresses at the boundary to traverse
the turbulent cascade all the way to dissipation.

     The mean vertical profile of the dissipation rate is, in rough terms,
a ``top-hat" stretching from $-3H$ to $+3H$.  It is not coincidental that
the dissipation rate drops sharply at the same place where magnetic pressure
begins to exceed gas pressure.  Dissipation is associated with sharp
gradients, particularly in the magnetic field, driven by turbulence.  At
higher altitudes, the comparative strength of the field keeps its structure
smooth and irons out sharp changes.  For many of the same reasons, the
vertical profile of the stress is very similar to that of the dissipation.

\begin{figure}
\epsscale{0.75}
\plotone{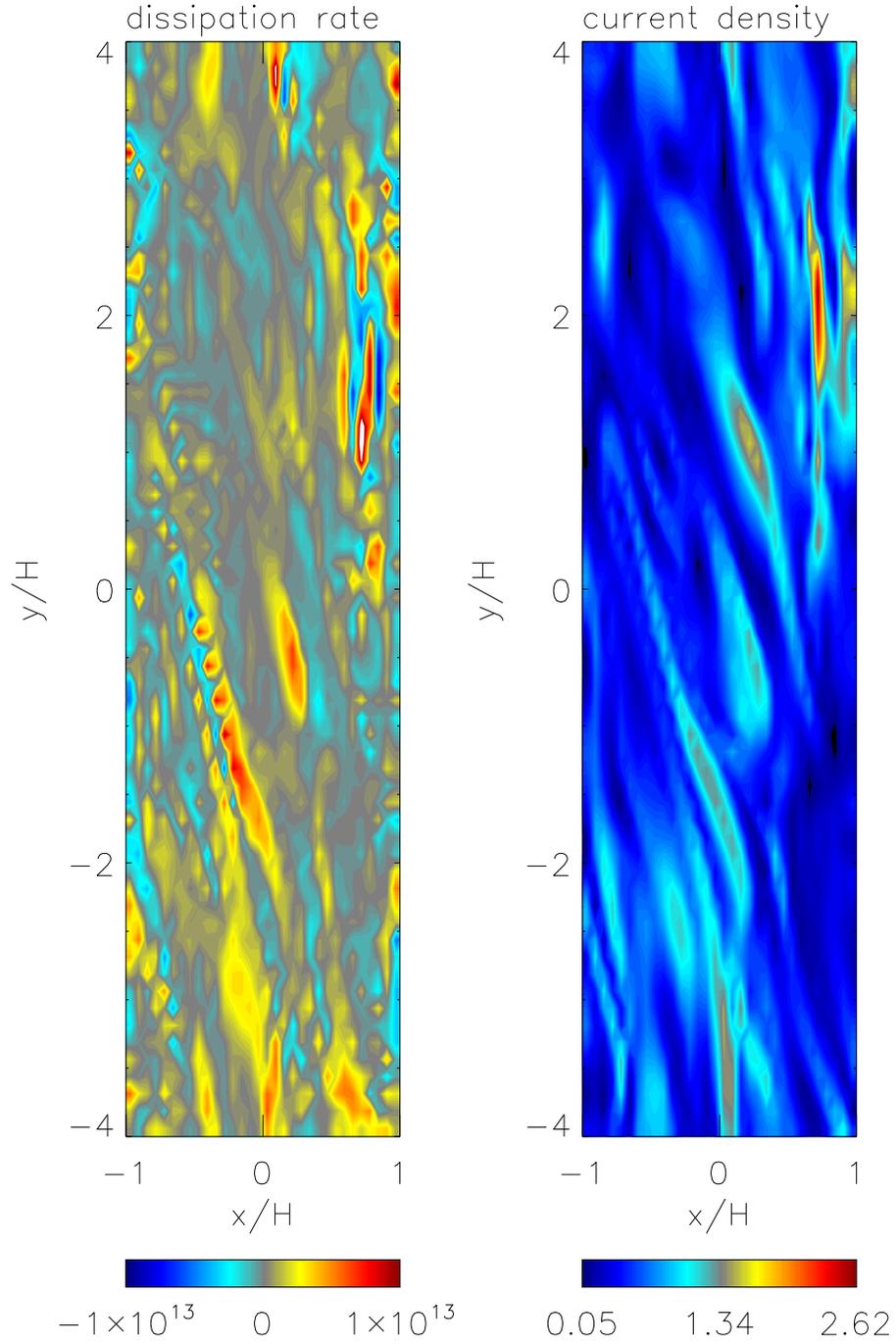}
\caption{A horizontal slice at $z=0$ at $t=35$~orbits.
Left panel: total dissipation rate (erg~cm$^{-3}$~orbit$^{-1}$).
Right panel: current density (esu~cm$^{-2}$~s$^{-1}$).
\label{fig:disscurrent}}
\end{figure}

\begin{figure}
\epsscale{1.0}
\plotone{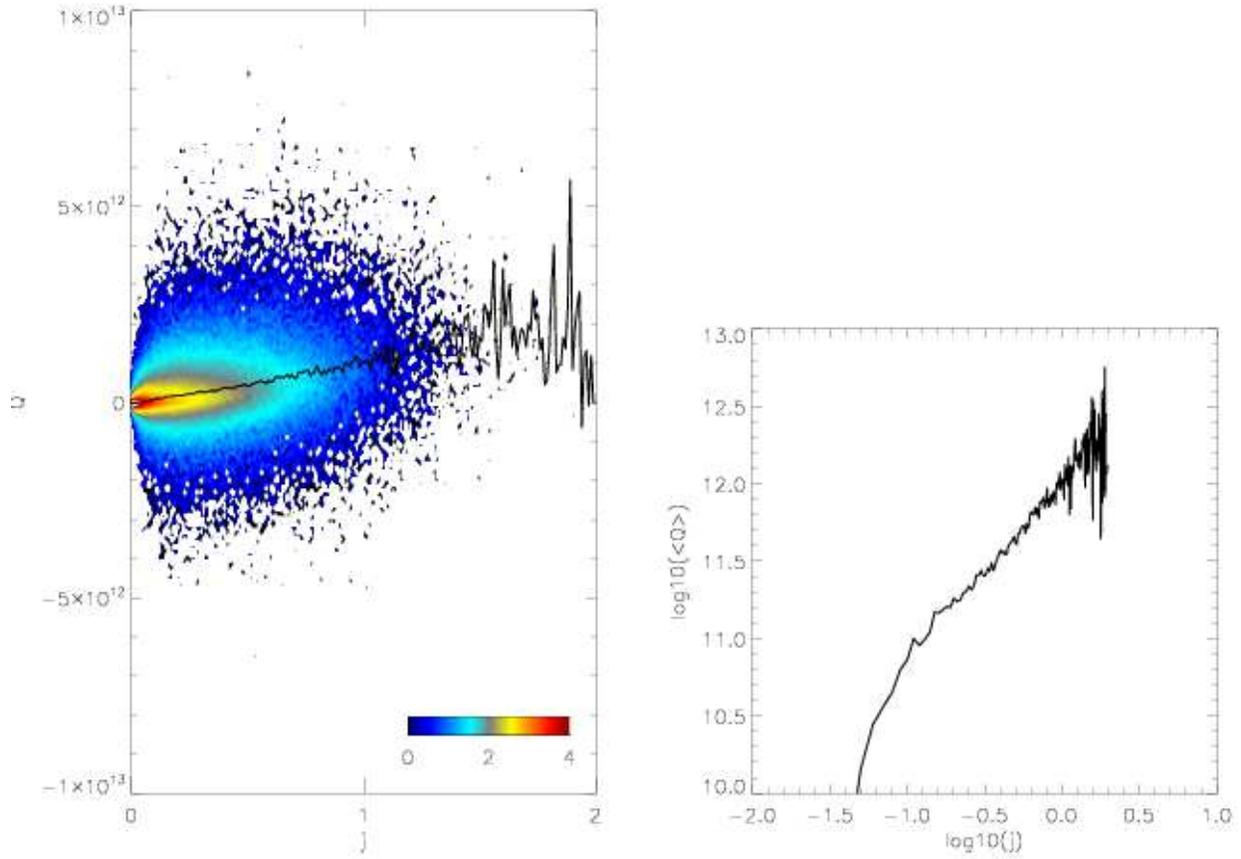}
\caption{The relation between current density and dissipation.  In the
main panel, the color indicates (on a logarithmic scale) the number of cells
having a given
level of current density and dissipation rate at time $t=40$~orbits.  The
curve running through it (reproduced in the inset) shows $\langle Q \rangle$,
the centroid of the dissipation distribution at fixed current density.
\label{fig:currdisscorr}}
\end{figure}

    A different view of the distribution of dissipation is also instructive
(Fig.~\ref{fig:disscurrent}).  Although low-level dissipation can occur
throughout the fluid due to numerical error, strong dissipation tends to be
correlated with intense current, i.e., regions of large $|\nabla \bm{B}|$.
A more quantitative view of the
association between current density and dissipation can be seen in
Figure~\ref{fig:currdisscorr}, where we plot the number of cells $N$
having a given dissipation rate $Q$ and current density $|\bm{J}|$ at
time $t=40$~orbits.  At small values of the current density,
a significant part of the local dissipation rate is due to random
numerical errors.  However, even when $|\bm{J}|$ is relatively small,
there is always a bias to positive dissipation.  Defining the
centroid of the distribution by
$\langle Q \rangle = \int \, dQ \, Q N(Q,|\bm{J}|)/\int \, dQ \, N(Q,|\bm{J}|)$,
we see that for the upper factor of 30 in its dynamic range,
$\langle Q \rangle$ is well-described by a power-law $\propto|\bm{J}|^{1.13}$.

\subsubsection{Cooling}\label{sect:cooling}

    The rate at which the heat content of the box equilibrates can
be parameterized by the cooling time.  In any accretion flow that
radiates efficiently, conservation of angular momentum and energy
require that $t_{\rm cool} \sim \alpha^{-1} P_{\rm orb}$.  One way to
evaluate it more quantitatively from the simulation data is to compute
the ratio of the total energy content of the box to the energy 
flux leaving through its top
and bottom surfaces.  Using this approach, we find a cooling time of 11.4
orbits.  In these terms, our simulation ran for a bit more than 5 cooling times,
with 4 of them well after the erasure of initial transients.

\begin{figure}
\epsscale{0.75}
\plotone{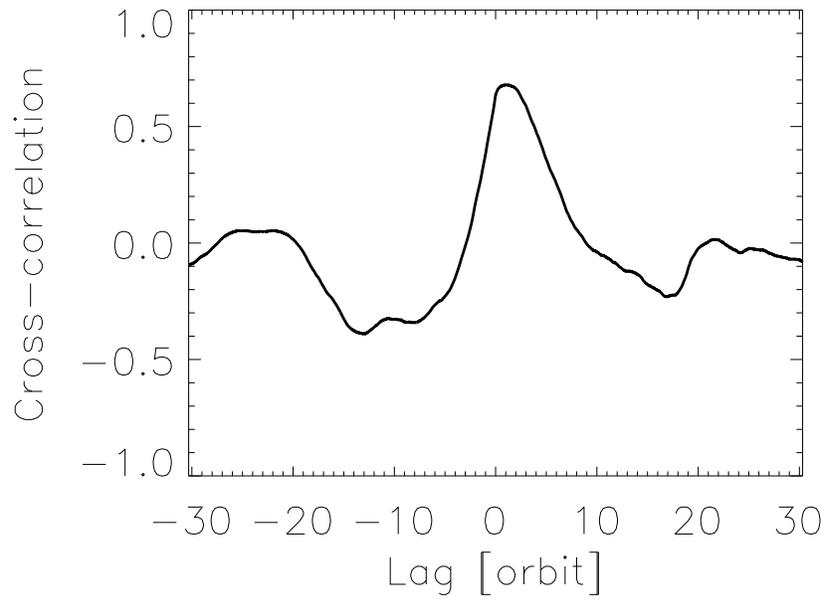}
\caption{Cross-correlation between volume-integrated dissipation rate and outgoing
flux.  
\label{fig:coolcrosscorr}}
\end{figure}

    However, another way to examine the delay between the creation of heat and
its loss is to study the cross-correlation between the volume-integrated 
dissipation and the flux out of the box.  That relationship is displayed in
Figure~\ref{fig:coolcrosscorr}.  As this figure shows, fluctuations in the flux
tend to follow those in the dissipation rate, but not by very much---although
there is an asymmetric tail toward positive lags, the cross-correlation falls 
to zero by +10 orbits, and its peak occurs in the range
0--2 orbits.  The reason that the cooling rate tracks the dissipation rate with
a lag almost an order of magnitude shorter than the nominal 
cooling time is that the variance in the dissipation rate is dominated by the region 
$2H \leq |z| \leq 3H$, where the cooling time is 10 to 30 times shorter than from
the midplane.   Much of the dissipation fluctuation power is on timescales
$\sim t_{\rm cool}$, but because the fluctuations take place in a region of
quick cooling, there is nonetheless little time-lag in imprinting them
on the radiation flux.

\begin{figure}
\epsscale{1.0}
\plottwo{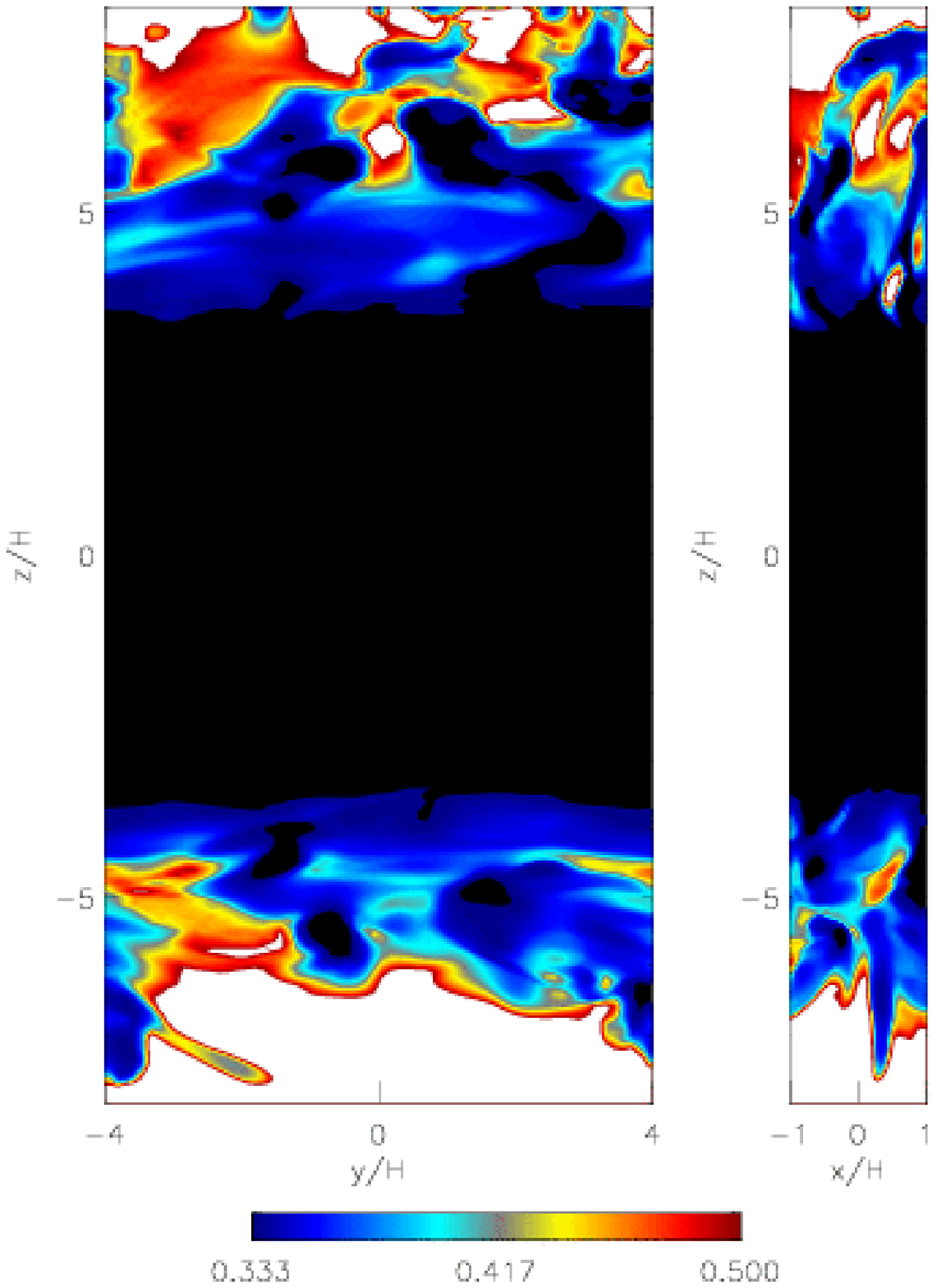}{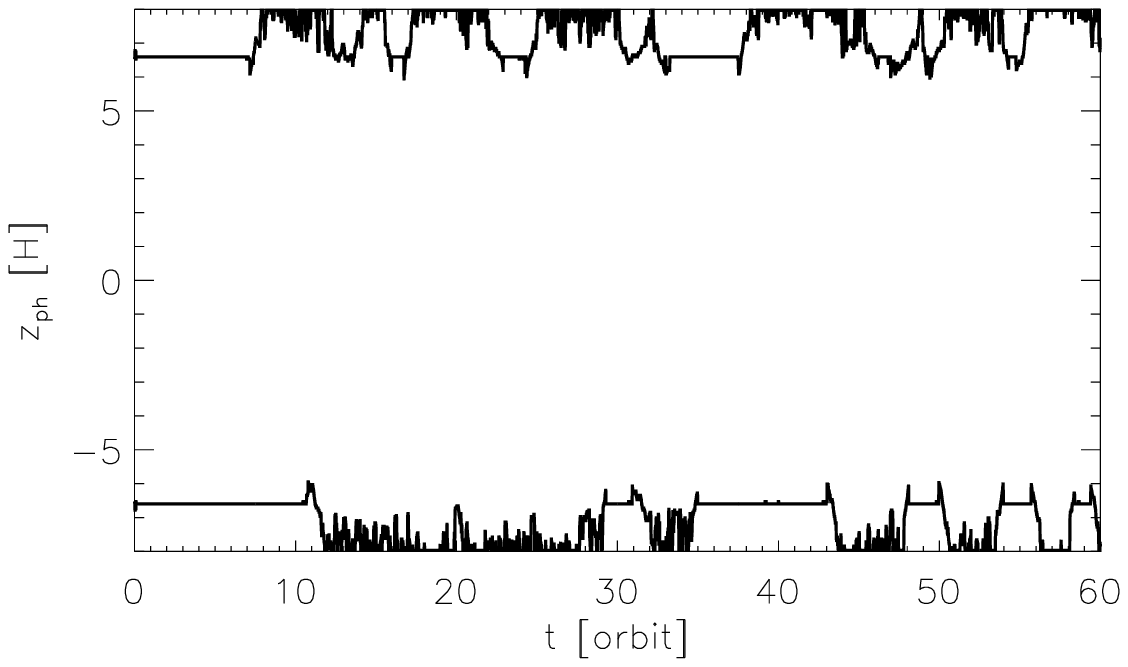}
\caption{Left panel: The Eddington factor in two vertical planes 
at $t=32$~orbits.  In all the white regions, the photons are
better described as free-streaming than diffusing.
Right panel: The horizontally-averaged photospheric altitude 
as a function of time.
\label{fig:photosphere}}
\end{figure}

   One of the goals of this effort was to compute the structure of the disk all
the way from the midplane out through the photosphere so that we could truly
say we were describing the entire flow of radiation energy through the disk.
Figure~\ref{fig:photosphere} shows the extent to which we succeeded in reaching
this goal.  In its left-hand panel, we show a snapshot of the Eddington 
factor $f$ as a function of position in the $x-z$ and $y-z$ planes.  
Deep within the optically-thick disk, the photon intensity is isotropic
and $f=1/3$; far above the photosphere, the photons become nearly
mono-directional and $f \rightarrow 1$.  We choose to define the photospheric
level as the place at which $f=1/2$.  By this standard, we see that the
photospheric surface at any given time is highly irregular.  Its height
above the midplane can be anywhere from $\simeq 6.5H$ to outside $\geq 8H$.
In fact, there are places where, as a function of $z$, the photon distribution
appears to pass back and forth with increasing altitude from nearly
free-streaming to more nearly diffusive.

    The snapshot shows roughness in the photospheric surface at a fixed time.
Some of the fluctuations are correlated in time, so that the horizontal
mean level of the photosphere also moves up and down over time (as portrayed
in the right-hand panel of Fig.~\ref{fig:photosphere}).  In this
horizontally-averaged sense, the photosphere generally varies between
$6.5H$ and $8H$ from the midplane, but there are occasional excursions both to
slightly lower and somewhat greater altitude.  The duration of these fluctuations
can vary from a small fraction of an orbit to 1--2 orbits.  We caution, however,
that our method of photosphere determination somewhat overestimates its height
from the midplane during times when the density in the outer portion of the box
is consistently at the floor.  Examples of such times can be identified in
the right-hand panel of Figure~\ref{fig:photosphere} as those periods during which
the nominal photospheric altitude is constant at $\simeq 6.5H$.  The fact that
the density floor is always the same accounts for the constancy of the estimated
photospheric height during these periods.

\subsection{Internal Fluctuations}
\begin{figure}
\epsscale{0.75}
\plotone{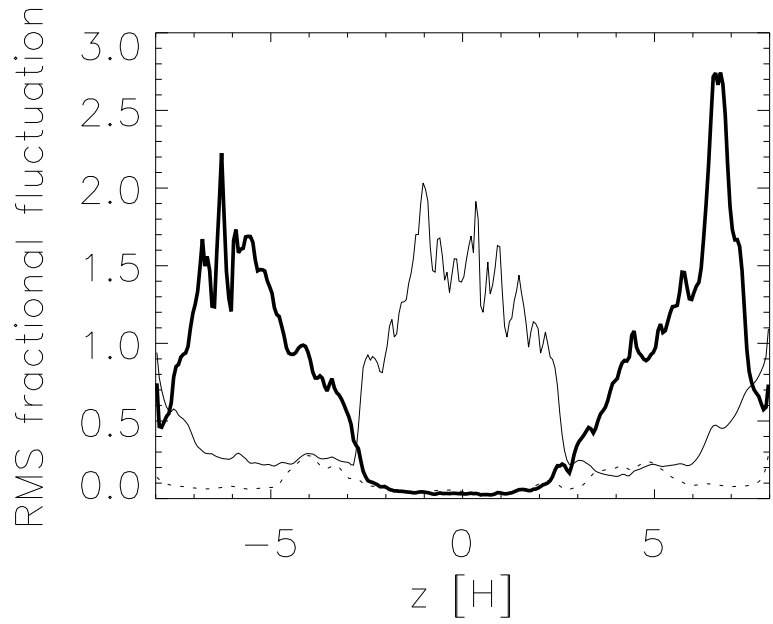}
\caption{RMS fractional fluctuation as a function of altitude at $t=25$~orbits.
Heavy solid line: gas density; thin solid line: magnetic field intensity;
dotted line: radiation energy.
\label{fig:flucts}}
\end{figure}

    MHD turbulence is, of course, at the heart of the dynamics in this system.
As a result, fluctuations are central.  We find that the character of these
fluctuations is a strong function of altitude from the midplane
(Fig.~\ref{fig:flucts}).  Because virtually every quantity has a strong
systematic dependence on $|z|$, we define all fluctuations as relative
to the mean at a given height.

\subsubsection{Gas density and magnetic intensity}
     The gas is almost incompressible near the midplane,
but the amplitude of density
fluctuations grows rapidly up to $|z| \simeq 6H$, and declines even more
sharply out to the top of the box.  Near $|z| \simeq 6H$, the rms fractional
fluctuation can be anywhere from $\simeq 1$--4 (at the time shown in
Fig.~\ref{fig:flucts} it is $\simeq 2.5$).  By contrast, the magnetic
field strength behaves in exactly
complementary fashion: with the exception of some likely spurious fluctuations
associated with the boundaries, it fluctuates the most in the densest part
of the disk, $|z| \leq 2.5H$.

    This behavior corresponds to the varying value of the plasma $\beta$
defined in terms of the sum of gas and radiation pressure.
Where $\beta$ is large (near the midplane), magnetic
effects cannot compress the gas; where it is small (roughly $2.5H \lesssim z 
\lesssim 7H$), magnetic forces can be very effective in driving gas compression
and rarefaction.  Large opacity guarantees such a strong dynamical coupling
between gas and radiation that their pressures sum for the purpose of resisting
magnetically-driven compression except on very small scales far from
the midplane.

\subsubsection{Velocity}\label{sec:velocity}

   Velocity fluctuations behave in a manner similar to density fluctuations.
In absolute terms, the largest random speeds are seen at high altitudes,
$|z| > 5H$.  In terms of the Mach number relative to the magnetosonic speed
(i.e., $c_s^2 \equiv [(5/3)p + (4/3)p_{\rm rad} + 2p_{\rm mag}]/\rho$),
the random motions are $\sim 0.3$ over most of the volume, but can sometimes
increase to $\sim 1$--2 near one or the other boundary.

    This behavior is qualitatively similar to what was seen in the simulations
of Miller \& Stone (2000).  They found that, measuring Mach number relative
to the gas sound speed alone, the rms fluid speed rose from $\simeq 0.15$
in the midplane to $\sim 1$ near the outer boundaries.  Although the
actual fluid speeds we see are considerably greater than the gas sound
speed, their Mach number with respect to the total sound speed is about
the same.

\subsubsection{Temperature}
\begin{figure}
\epsscale{0.5}
\plotone{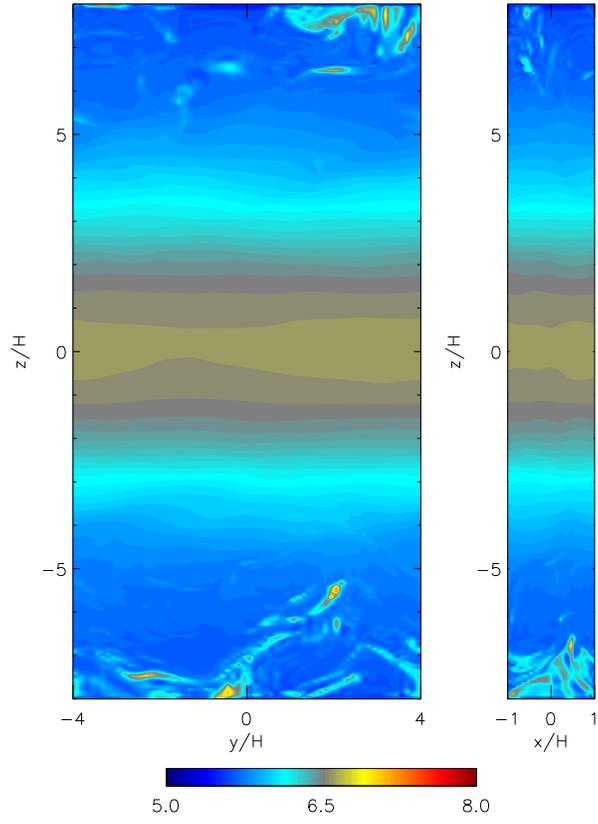}
\caption{Temperature on a logarithmic scale in two vertical planes at
time $t=25$~orbits.
\label{fig:tempflucts}}
\end{figure}

   In the disk body, the temperature is generally quite uniform at any given
altitude.  However, in the region above $5H$, at any given time there is
generally a few percent of the volume in which the
temperature is elevated by a factor of 5 or so above the mean
(Fig.~\ref{fig:tempflucts}).  Strong negative fluctuations are quite rare.
Most often these regions of elevated temperature are roughly round,
anywhere from $0.1H$ to $0.5H$ in diameter (so the larger ones are
well-resolved by our grid), but occasionally they can be sheet-like.

\subsubsection{Top-bottom asymmetry}\label{sec:asymm}
\begin{figure}
\epsscale{0.75}
\plotone{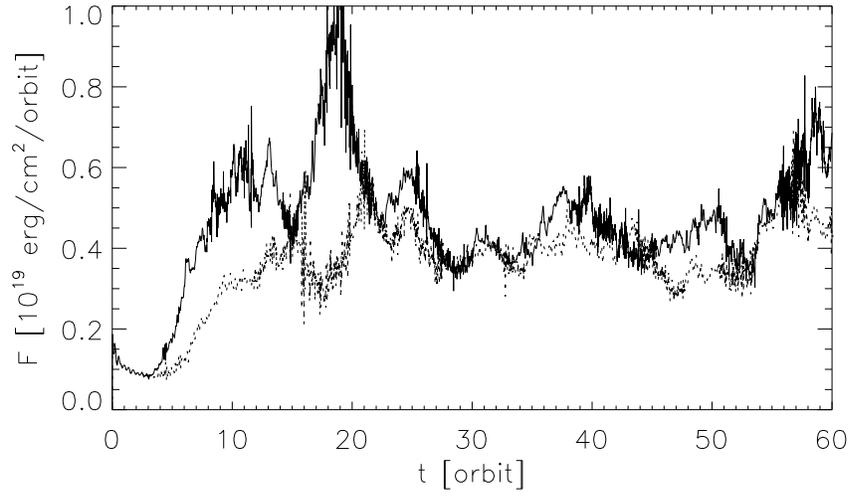}
\caption{Radiation flux leaving the top (solid curve) and bottom (dotted curve)
surfaces as functions of time.
\label{fig:topbotflux}}
\end{figure}

   A different sort of fluctuation may be seen in space-time plots.
Although in the long-run the top and bottom halves of the disk behave
symmetrically, as shown
in Figure~\ref{fig:magspacetime}, the magnetic field can be persistently
stronger on one side of the midplane than on the other.  Because
energy dissipation is largely magnetic, during periods of asymmetry
in field strength there is generally a corresponding asymmetry in dissipation
rate.  Moreover, the fact that much of the dissipation takes place several
scale-heights from the midplane means that an asymmetry in dissipation is
mirrored in a dissipation in flux.  For periods as long as 5--10 orbits,
the radiation flux through the top or bottom surface can be as much as 2--3
times as much as through the opposite surface (Fig.~\ref{fig:topbotflux}), 
although through most of the simulation the contrast was smaller, 
typically $\sim$ 20\%.

\subsection{Magnetic fieldline structure}
\begin{figure}
\epsscale{0.5}
\plotone{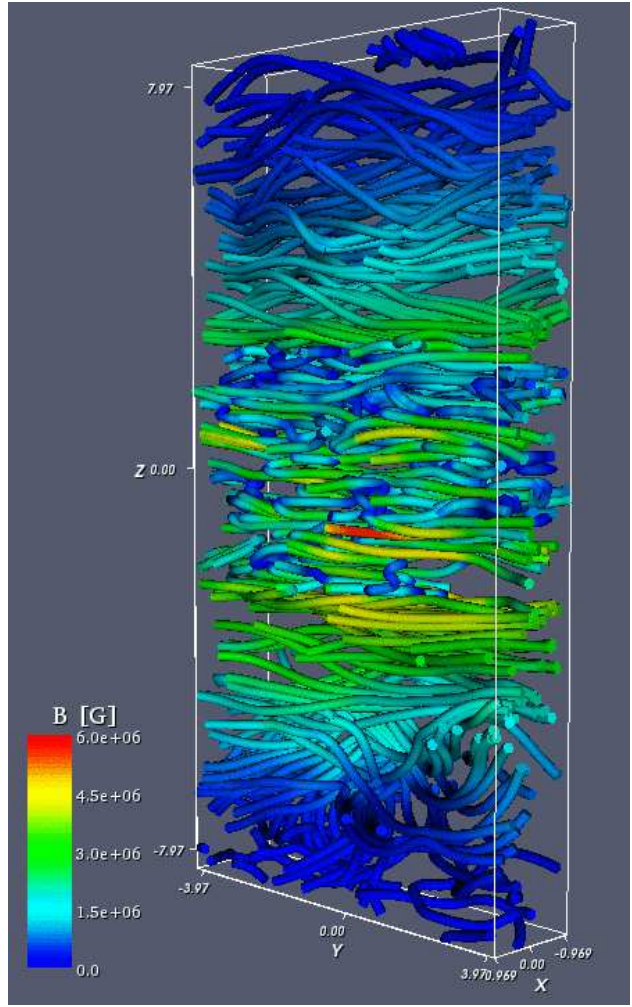}
\caption{Magnetic field structure at $t=25$ orbits. Color represents
the magnitude of the magnetic field in Gauss.  Starting points of
field-line integrations are randomly distributed in the box, and thus
the density of field-lines is not proportional to the magnitude.}
\label{fig:fieldline}
\end{figure}

A view of the magnetic field structure in the steady state is shown in
Figure \ref{fig:fieldline}.  The field-lines are predominantly azimuthal
throughout the box, but subtle differences can be distinguished as a function
of altitude.  Near the midplane ($|z|<2H$), the field-lines are tangled on
short scales ($\sim 0.1H$), corresponding to the MHD turbulence driven
by the MRI.  At mid-altitudes ($2H<|z|<5H$), the field-lines are
generally fairly smooth.
In the high-altitude zone ($5H<|z|$), the field-lines
get disordered again, and the scale of wiggles is larger ($\sim H$) compared
to that near the midplane.  This is because 
random fluid motions are powerful enough (see section \ref{sec:velocity})
to substantially perturb fieldlines.

\section{Discussion}
\subsection{What is a corona?}
   Perhaps the most striking structural feature apparent in this simulation
is the $|z| > 3H$ region, where most of the support against gravity is from
magnetic pressure.  On the basis of its low plasma $\beta$, it might be
described as a corona, but it is a very unusual corona in that it lies
well below the photosphere.  Because of this magnetic support, the top surface
of the disk (if ``top" is defined as the photosphere), is considerably higher
above the midplane than where the disk surface is conventionally estimated to be.
An isothermal Gaussian model, for example, places the photosphere at
merely $3H$, rather than $>7H$ as we find.
Although a similarly magnetically-supported region existed in the 
\citet{Miller & Stone(2000)} simulations, 
it went unremarked because they did not evaluate the optical depth.

    The nature of this region also contradicts another frequently-held view:
that magnetically-dominated layers at 
high altitude should be the site of intense
heating, so intense that the temperature can be raised high enough to support
hard X-ray emission.  This idea was given indirect support by the
\citet{Miller & Stone(2000)} finding of substantial vertical Poynting flux
in the region a few scale-heights from the midplane, although they could not
test it directly because they assumed an isothermal equation of state.
We find rather less Poynting flux than did they, and little
heat dissipation in the magnetically-dominated region.  

   The absence of dissipation in the ``corona" is most likely a result of
the very fact that this zone is magnetically-dominated: the field smooths
itself.  We do see a narrow
zone of high dissipation rate immediately inside the outer boundaries, and
the dissipation there does raise the temperature by about a factor of two,
but at least part (and maybe all) of this heating near the boundary
is likely to be a numerical artifact.  We also observe transient localized
regions of very high temperature in this ``corona" well below the outer 
boundary, but these are neither hot enough nor common enough to count
for much in terms of global properties.

   These results raise the question of where the observed hard X-rays are
made.  Perhaps more promising sites for their production can be found at
smaller radii.  Where radiation pressure dominates gas pressure at all 
altitudes, the disk's internal vertical profile may be very different.
In addition, as demonstrated in \citet{Hirose et al.(2004)}, there can
be extremely high current-density regions at radii comparable to or
smaller than the marginally stable orbit and
well away from the equatorial plane.  Hirose et al. speculated that these,
too, may be loci of hard X-ray production.

\subsection{Observed radiation}

    In comparison to the commonly-assumed Gaussian density profile,
the extended magnetically-supported atmosphere found in our simulation
can be expected to alter the locally-emitted spectrum in significant ways.
Specifically, the very steep density gradient at the photosphere predicted
by the Gaussian model entails a large density there.  In
the lower density environment we find near the photosphere, larger departures
from LTE might be expected, and therefore stronger non-thermal features
in the emergent spectrum.  A symptom that something of this sort may
happen can be seen in the band, a few scale-heights thick, in which
the gas temperature rises well above the local radiation temperature.

    In the great majority of atmosphere calculations, whether in the context of
stars or accretion disks, it is assumed that the photospheric region may be
approximated as time-steady and having 1-d slab geometry.
Our results regarding
both the location of the photosphere and temperature fluctuations in the
upper disk call both approximations into question.  However, the nature of
radiation transfer may ameliorate both problems.  Because the photon
diffusion time is very short compared to the local dynamical time in the
upper atmosphere of the disk, the radiation can respond very quickly
to varying conditions.  The nature of its response is for
diffusion to smooth out irregularities; in fact, there is far less spatial
variation in horizontal planes in the radiation intensity than in the
gas density or temperature or magnetic field strength 
(Fig. \ref{fig:flucts}).  In addition,
the characteristic temperature fluctuation takes the form of a spot of high
temperature, but because the opacity diminishes
with increasing temperature, these spots largely decouple from the radiation.

    Another virtually universal assumption made in previous work on accretion
disks is that their top and bottom surfaces
should have equal surface brightness.  Yet we find that the
magnetic field intensity, and therefore the dissipation rate, and consequently
the emergent flux, can be asymmetric with respect to the midplane with
a consistent sense of contrast for as long as 5--10 orbits at a stretch.
Because we are simulating a shearing-box and not an entire disk, we have no
way of knowing how well correlated such asymmetries might be in radius.
If there is significant coherence in these fluctuations over radial extents
comparable to the local radius, the luminosity of the disk as measured by
any single observer could be wrong by factors of a few.

    It is likewise frequently asserted that the radiation from a disk 
surface cannot vary in any substantial way on timescales shorter than a cooling
time.  We have found, however, that because so much of
the dissipation variation takes
place at comparatively high altitude, it is entirely possible for the emergent
flux to vary as much as an order of magnitude faster than the global 
equilibration cooling time.

\subsection{Scaling with central mass}
    Most properties of accretion disks depend only on quantities such as the
accretion rate in Eddington units $\dot m$ and the radius in gravitational
radii from which the mass of the central object has already
been scaled out.  However, the temperature is different.  For fixed
$\dot m$ and $r/r_g$, the temperature is $\propto M^{-1/4}$.  The most
significant effects that this is likely to have on the properties discussed
here all result from the dependence of the opacity on temperature.

    As we mentioned earlier, we chose a relatively small central mass
($6.6 M_{\odot}$) so that sources of opacity other than free-free
absorption and Thomson opacity would be comparatively unimportant.
However, if the mass were $\sim 10^5$
times greater, with $\dot m$ and $r/r_g$ fixed at the values of our
simulation (0.1 and 300), the surface temperature would
fall to $\simeq 4 \times 10^4$~K.
At this temperature, there would be substantial opacity due to resonance
lines in medium-$Z$ elements.  It is possible that the scattering opacity
would then be so large that the vertical radiation force could overcome
the vertical component of gravity even at high altitude (cf. Proga
\& Kallman 2004).  In this circumstance,
the vertical structure of the disk would undoubtedly be quite different.

    In a similar vein, at still higher masses ($\sim 10^8 M_{\odot}$, the
domain of AGNs), the surface temperature would be $\simeq 1 \times 10^4$~K,
where ionization transitions can lead to pulsational instabilities.

\section{Conclusions}

    We have shown that it is now possible to track in some detail
the path of energy within accretion disks, from turbulent dissipation
to heat to diffusing photons.  When we do so, we find that some conclusions
about disk structure drawn on simpler thermodynamic models (e.g., isothermal
equations of state) are supported, but not all.  For example, the existence
of an extended magnetically-supported atmosphere was already predicted
by the isothermal models.  What they could not do, however, because
they did not compute optical depths, was to discover that a large part
of this magnetically-supported atmosphere can lie below the photosphere.

    Our model for the dissipation of turbulence will certainly not be
the final word on this subject---we do not have the resolution to
capture the full inertial range of MHD turbulence driven by the MRI, nor
do we compute its
genuine microphysics.  However, locating dissipation
where the gradients are so sharp as to cause numerical losses of
magnetic and kinetic energy is a very plausible supposition.  On the
basis of this method, we have found that the time-averaged dissipation profile
is roughly flat in terms of rate per unit volume within roughly
$\pm 3H$ of the midplane, but drops sharply outside that height,
with very little dissipation in the magnetically-supported (but
optically-thick) ``corona".  Over timescales that are presumably
short compared to the lifetime of the disk, but comparable to
or even somewhat greater than the local cooling time, there can
be sizable departures from this mean profile, especially in the
sense of an asymmetry between the top and bottom halves of the disk.

     We have also found that this spread in the dissipation permits
significant parts of it to be found in places where the photon
diffusion time to the surface is relatively short.  Consequently,
there can be significant fluctuations in the outgoing flux that
track fluctuations in the disk heating rate with a lag that is
an order of magnitude shorter than the nominal cooling time.

     Lastly, the detailed structure we have computed suggests that
the emergent spectrum from the gas-dominated portions of accretion
disks may deviate significantly from blackbody.  They are likely
to have extended atmospheres in which the gas temperature differs
from the radiation temperature and the densities are low enough
that other breakdowns in the details of LTE may occur.  In addition,
the photospheric surface itself may have both a complicated topology
and interesting fluctuations, leading to additional non-blackbody
effects.

\acknowledgments

     We would like to thank Neal Turner, who contributed so much
to the previous version of our simulation code.

     This work was partially supported by NSF Grants AST-0205806 and AST-0313031
(SH, JHK), and by NSF grant AST-0413788 and Princeton University (JMS).

\appendix
\section{Implementation of Total Energy Conservation}\label{app:cnsv}

\subsection{Operator-splitting}
The numerical scheme of the ZEUS code, upon which our code is built, consists
of the following seven steps:
\begin{enumerate}
\item source step (inertial and gravitational part)
\begin{eqnarray}
&&\frac{\partial \bm{v}}{\partial t}=-2\rho\left(\Omega\hat{\bm{z}}\right)\times\bm{v}+3\rho\Omega^2x\hat{\bm{x}}-\rho\Omega^2z\hat{\bm{z}}
\end{eqnarray}
\item source step (radiation part)
\begin{eqnarray}
&&\frac{\partial \bm{v}}{\partial t}=-\frac{\chi\rho}{c}\bm{F}\\
&&\frac{\partial e}{\partial t}=-\left(4\pi B-cE\right)\kappa\rho\\
&&\frac{\partial E}{\partial t}=\left(4\pi B-cE\right)\kappa\rho-\mathsf{P}:\nabla\bm{v}
\end{eqnarray}
\item radiation diffusion
\begin{eqnarray}
&&\frac{\partial E}{\partial t}+\nabla\cdot\bm{F}=0
\end{eqnarray}
\item gas pressure gradient
\begin{eqnarray}
&&\frac{\partial \bm{v}}{\partial t}=-\frac{\nabla p}{\rho}\\
&&\frac{\partial e}{\partial t}=-p\left(\nabla\cdot\bm{v}\right)\label{eq:tot4}
\end{eqnarray}
\item artificial viscosity
\begin{eqnarray}
&&\frac{\partial \bm{v}}{\partial t}=-\frac{\nabla q}{\rho}\\
&&\frac{\partial e}{\partial t}=-q\left(\nabla\cdot\bm{v}\right)\label{eq:tot5}
\end{eqnarray}
\item transport step
\begin{eqnarray}
&&\frac{\partial\rho}{\partial t}+\nabla\cdot\left(\rho\bm{v}\right)=0\\
&&\frac{\partial \rho\bm{v}}{\partial t}+\nabla\cdot\left(\rho\bm{v}\bm{v}\right)=0\\
&&\frac{\partial E}{\partial t}+\nabla\cdot\left(E\bm{v}\right)=0\\
&&\frac{\partial e}{\partial t}+\nabla\cdot\left(e\bm{v}\right)=0\label{eq:tot6}
\end{eqnarray}
\item magnetic part
\begin{eqnarray}
&&\frac{\partial \bm{v}}{\partial t}=\frac{1}{\rho}\bm{j}\times\bm{B}\label{eq:vel7}\\
&&\frac{\partial\bm{B}}{\partial t}+\nabla\times\bm{E}=0\label{eq:mag7}\\
&&\frac{\partial e}{\partial t}= 0 
\label{eq:int7}
\end{eqnarray}
\end{enumerate}

To transform this into an energy-conserving numerical scheme, several changes
are required.  \citet{Turner et al.(2003)} began this transformation by
altering step 7 to be the solution of a total energy equation in the 
conservative form
\begin{eqnarray}
&&\frac{\partial}{\partial t}\left(\frac{1}{2}\rho\bm{v}^2+e+\frac{1}{8\pi}\bm{B}^2\right)+\nabla\cdot\left(\frac{1}{4\pi}\bm{E}\times\bm{B}\right)=0 .
\end{eqnarray} 
They then updated the internal energy according to
\begin{equation}
e^*=\left(\rho\frac{\bm{v}^2}{2}+e+\frac{\bm{B}^2}{8\pi}\right)^*
     -\left(\rho^*\frac{{\bm{v}^*}^2}{2}+\frac{{\bm{B}^*}^2}{8\pi}\right),
\end{equation}
where the asterisk variables denote those updated in the step.  In
this way, any kinetic or magnetic energy that might have been lost numerically
during this step is instead captured into internal energy because 
the total energy $\rho\bm{v}^2/2+e+\bm{B}^2/(8\pi)$ is conserved.

We have extended this approach to steps 4, 5, and 6, in order to
capture in the same way any kinetic energy that would otherwise be lost.
In each of these steps, the internal energy equation is replaced by
the corresponding total energy equation:
\begin{enumerate}
\item[4.] gas pressure gradient
\begin{eqnarray}
&&\frac{\partial}{\partial t}\left(\frac{1}{2}\rho\bm{v}^2+e\right)+\nabla\cdot\left(p\bm{v}\right)=0
\end{eqnarray}
\item[5.] artificial viscosity
\begin{eqnarray}
&&\frac{\partial}{\partial t}\left(\frac{1}{2}\rho\bm{v}^2+e\right)+\nabla\cdot\left(q\bm{v}\right)=0
\end{eqnarray}
\item[6.] transport step
\begin{eqnarray}
&&\frac{\partial}{\partial t}\left(\frac{1}{2}\rho\bm{v}^2+e\right)+\nabla\cdot\left\{\left(\frac{1}{2}\rho\bm{v}^2+e\right)\bm{v}\right\}=0
\end{eqnarray}
\end{enumerate}
After each of these steps, the internal energy is updated by
\begin{equation}
e^*=\left(\rho\frac{\bm{v}^2}{2}+e\right)^*-\left(\rho^*\frac{{\bm{v}^*}^2}{2}\right).
\end{equation}
Thus, the kinetic and magnetic energies that in the original ZEUS code
would be lost numerically  are completely captured as internal energy
during steps 4 to 7 in our scheme.  The net result of all these steps
is represented by the symbol $Q$ in Equation~\ref{eq:eint}.  We show
how it is explicitly computed in Appendix~\ref{app:diss}.


\subsection{Dissipation rate}\label{app:diss}

The dependent variables $e$, $E$, $\rho$, $\bm{B}$, and $\bm{v}$
suffice to determine the state of the simulation.  However, we
are interested in the dissipation as a function of time and
position as a fundamental diagnostic of the thermal physics
in these disks.  Because of the way we update the internal
energy, to record the dissipation rate requires some special
effort.

   The total dissipation rate receives contributions in all the
steps from 4 through 7.
To compute their individual rates, we retain the internal 
energy equation and solve it in parallel with the total energy equation.
At each step, we must distinguish between the internal energy as it
evolves and the internal energy that would have resulted from
adiabatic changes alone.  For the purposes of this discussion,
we convey this distinction by defining $e$ as the internal energy
before an update step, $e^*$ as the internal energy
after update by the total energy equation, and $\tilde e^*$
as the internal energy after an update by the internal
energy equation alone.

The artificial viscosity dissipation rate 
in step 5 can be computed as $(\tilde e^*-e)/\Delta t$ because the RHS of 
the internal energy equation in this step is nothing more than this
dissipation rate.  On the other hand, 
the numerical dissipation rate in each of steps 4 through 7
can be evaluated as $(e^*-\tilde e^*)/\Delta t$ because the difference 
$e^*-\tilde e^*$ corresponds to the numerically-lost kinetic and magnetic
energies captured as internal energy.

The total dissipation rate in the simulation can then be written as 
the sum of all these rates:
\begin{equation}
  Q=\left.{\frac{\tilde e^*-e}{\Delta t}}\right|_{\rm step=5} +
    \sum_{\rm step=4,5,6,7}\frac{e^*-\tilde e^*}{\Delta t}.
\end{equation}
Note that the artificial viscosity dissipation rate $(\tilde e^*-e)/\Delta t$ 
is always positive by definition, whereas a numerical dissipation 
rate $(e^*-\tilde e^*)/\Delta t$ can be negative due to numerical errors.

   Because our evaluation of the dissipation rate intrinsically involves
capturing numerical errors that are likely to depend strongly on the
grid-scale relative to the scale of physical gradients, one might
fairly ask whether our results are sensitive to resolution.  To
answer this question, we have run a shorter simulation with a grid
having twice the resolution in the $y$-direction.  The initial condition
for this resolution-testing simulation was the state of the main
simulation at $t = 35$~orbits, and it ran for 1 full orbit.  To
distinguish the two, we call our primary simulation the ``standard".

\begin{figure}
\epsscale{1.2}
\plottwo{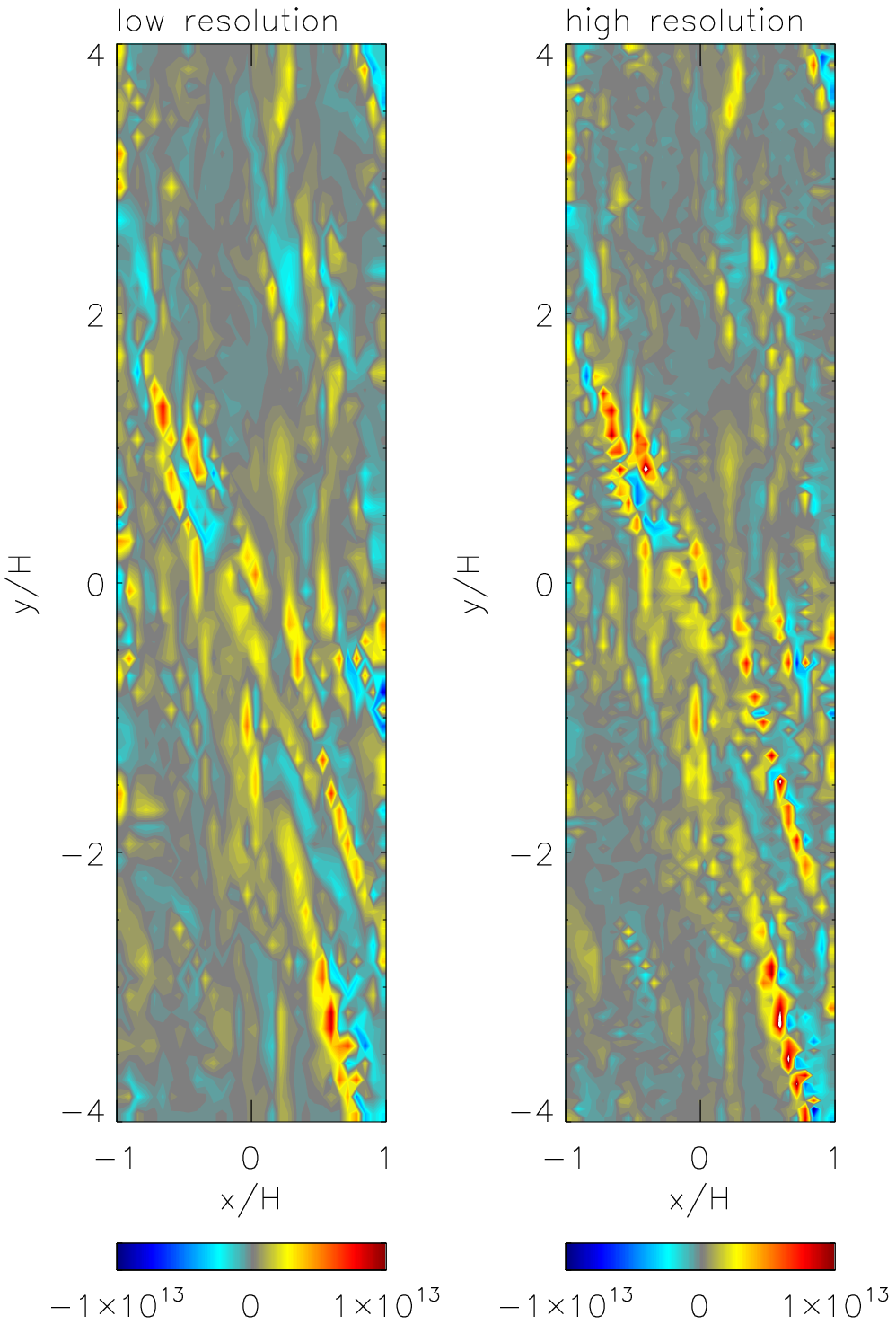}{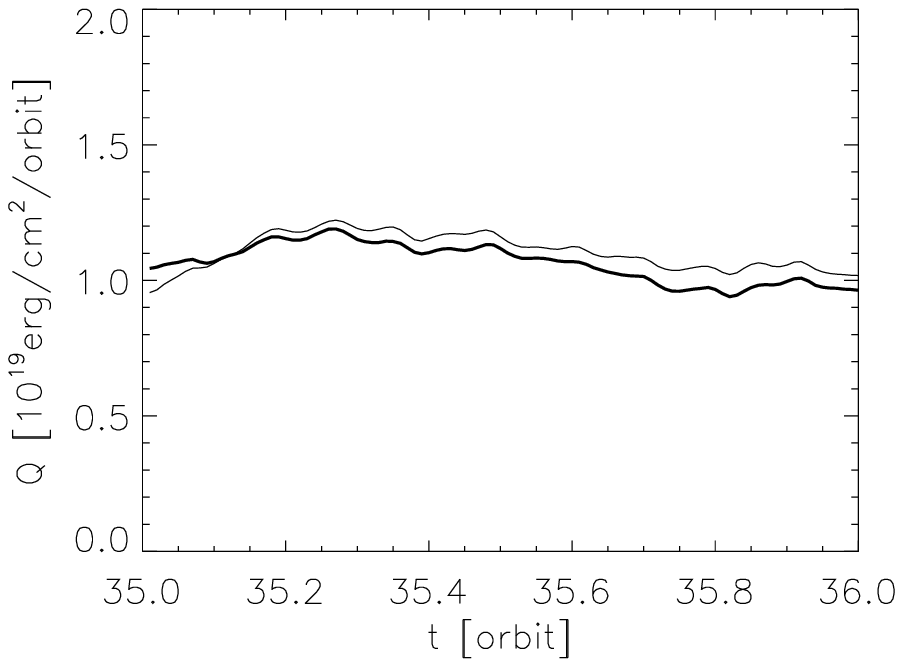}
\caption{Left panel: Dissipation rate per unit volume at $z=0$ 
and $t=35.1$~orbits in the standard simulation (left-hand image)
and the high-resolution simulation (right-hand image).  Right panel:
Volume-integrated dissipation rate for the standard simulation
(solid curve) and high-resolution simulation (dotted curve).
\label{fig:resolutioncheck}}
\end{figure}

    For times up to a few tenths of an orbit after its start, the
distribution of dissipation in the higher-resolution simulation remained
very similar to that in the standard one with the
exception that peaks in the dissipation were more sharply-defined
(Fig.~\ref{fig:resolutioncheck}).
After a few tenths of an orbit, the chaotic character
of the turbulent dynamics leads to a divergence of the detailed structure
of the flow as compared to the standard simulation.  Nonetheless, the
time-history of the volume-integrated dissipation rate in the
higher-resolution simulation tracks closely that of the standard for
the duration of the test, deviating by only a few percent at most over
an orbital period.  Thus, we conclude that if the magnetic structure
is independently fixed in place, our numerical scheme locates the
dissipation more or less as well as it can, given the constraints posed
by the actual resolution employed.
Over timescales long enough to move the specific places where dissipation
happens, the fact that the volume-integrated dissipation rate hardly
changes in this limited resolution test means that we have no immediate
grounds for concern about sensitivity of our results to our resolution
scale.

\subsection{Artificial energy injection}\label{app:arti}
Three limiting values must be placed on quantities in order to preserve
stability and to keep the time-step from becoming prohibitively small.
These are: a density floor, an energy floor, and a velocity cap.
The density floor and the velocity cap are applied only after step 7,
but the energy floor is applied in each of 
steps 4 to 7.  Here we describe how they work 
and evaluate energy injection rates associated with each of them.

(a) density floor:
An extremely low density would cause a very short time-step because
$\Delta t$ must be $< \Delta x /c_{\rm ms}$.
If the density were to go negative, it could cause the code to halt.
To avoid either of these problems,
the density $\rho$ is set to the floor value $\rho_{\rm floor}$ 
when $\rho<\rho_{\rm floor}$.  For this simulation, we set
$\rho_{\rm floor}$ to $10^{-5}$ times 
the initial density at the midplane. The energy injection rate due to the 
density floor is 
\begin{equation}
  S_1\equiv \cases{
    (\rho_{\rm floor}\bm{v}^2/2-\rho\bm{v}^2/2)/\Delta t & (if $\rho<\rho_{\rm floor}$)\cr
    0                                                    & (otherwise).
}
\end{equation}

(b) energy floor:
The internal energy computed by the total energy equation, $e^*$, can become
extremely small or even negative due to numerical errors. 
Simply applying a floor of small value does 
not work because it can lead to an unphysically large free-free opacity. 
Therefore, we employ $\tilde e^*$ instead of $e^*$ when $e^*<f\tilde e^*$; 
the fudge factor $f=0.5$ in this simulation. 
The energy injection rate is then
\begin{equation}
  S_2\equiv \cases{ 
    \sum_{\rm step=4}^{7}(\tilde e^*-e^*)/\Delta t &(if $e^*<f\tilde e^*$)\cr
    0                                              &(otherwise).             \cr
  }
\end{equation}

(c) velocity cap:
Two different limits are placed on the velocity: 
(1) To avoid inflows through the top and bottom boundaries, 
the $z$-component of the velocity at the top (bottom) 
boundary surfaces is forced to zero when it is negative (positive). 
(2) To avoid unusually large speeds due to numerical errors, 
a cap is applied to the magnitude of each component of the velocity; 
the value of the velocity cap is set to $10(3/2)\Omega L_x$, where
$L_x$ is the full-width of the box in the $x$-direction.
The energy injection rate due to these corrections is
\begin{equation}
  S_3\equiv \cases{
    (\rho\bm{v}_{\rm corrected}^2/2-\rho\bm{v}^2/2)/\Delta t & (if $\bm{v}$ is corrected)\cr
    0                                                       & (otherwise),
}
\end{equation}
which is always negative by definition.

   Integrated over the 60 orbits of the simulation, each of these is quite
small, but the total amounts of energy injection due to the energy floor
and velocity cap are rather smaller than that due to the density floor.
Relative to the total dissipated energy, the density floor adds 0.9\%, the
energy floor subtracts 0.06\%, and the velocity cap subtracts 0.05\%.

\subsection{Testing energy conservation}
\begin{figure}
\epsscale{0.75}
\plotone{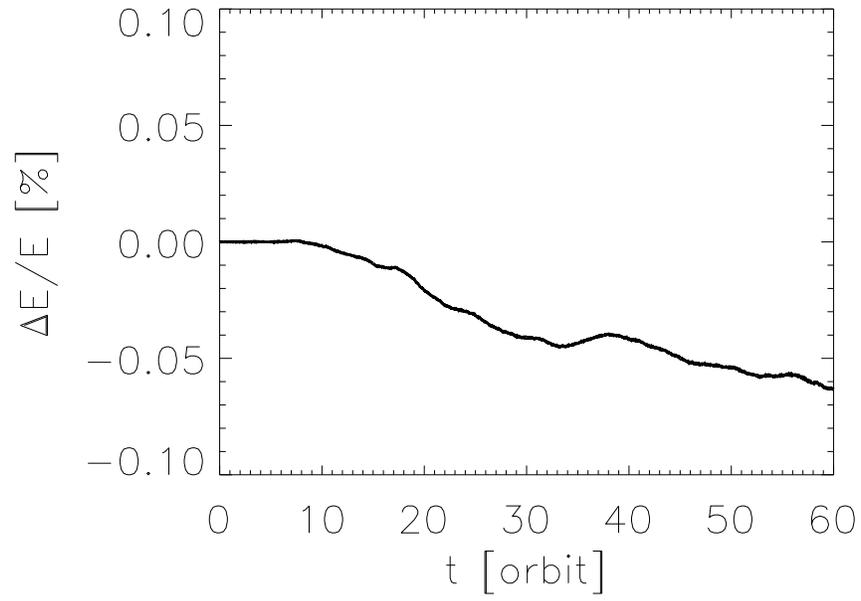}
\caption{The energy conservation criterion described in text as a function
of time through the simulation.
\label{fig:econserve}}
\end{figure}

Total energy conservation in our system can be written as follows.
\begin{eqnarray}
&&\frac{\partial \mathcal{E}}{\partial t}+\nabla\cdot\bm{\mathcal{F}}=\mathcal{S}\label{eq:totcnsv}\\
&&\quad \mathcal{E}\equiv\frac{1}{2}\rho\bm{v}^2+e+\frac{1}{2}\bm{B}^2+E\\
&&\quad \bm{\mathcal{F}}\equiv\left(\frac{1}{2}\rho\bm{v}^2+e\right)\bm{v}+\left(p+q\right)\bm{v}+\bm{E}\times\bm{B}+E\bm{v}+\bm{F}\\
&&\quad \mathcal{S}\equiv\left\{-2\rho\left(\Omega\hat{\bm{z}}\right)\times\bm{v}+3\rho\Omega^2x\hat{\bm{x}}-\rho\Omega^2z\hat{\bm{z}}\right\}\cdot\bm{v}+\left(\frac{\chi\rho}{c}\bm{F}\right)\cdot\bm{v}-\mathsf{P}:\nabla\bm{v}\nonumber\\
&&\qquad +S_1+S_2+S_3
\end{eqnarray}
Integrating equation (\ref{eq:totcnsv}) over time and volume, we have
\begin{equation}
\bar\mathcal{E}(t)=\bar\mathcal{E}(0)+\int_{0}^t \, dt^{\prime}
\left\{-\bar\mathcal{F}(t^{\prime})+\bar\mathcal{S}(t^{\prime})\right\},
\end{equation}
where $\bar\mathcal{E}$, $\bar\mathcal{F}$, and $\bar\mathcal{S}$ are, 
respectively, volume integrals of $\mathcal{E}$, $\nabla\cdot\bm{\mathcal{F}}$,
and $\mathcal{S}$. To evaluate how well total energy is conserved 
in our numerical code, we computed a relative error 
$({\rm LHS}-{\rm RHS})/{\rm LHS}$ 
as a function of time, which is shown in Figure~\ref{fig:econserve}.
We see that the absolute value of the error stays lower than $0.1\%$
over a span of 60 orbits.

\section{Boundary Condition for Radiation Diffusion Equation}
\label{app:ebndc}
The top and bottom boundary conditions for the radiation
diffusion equation 
(step 3 in Appendix \ref{app:cnsv}) are determined by a requirement that 
the diffusion flux $F_z=-DdE/dz$ across 
the boundary is equal to that across the adjacent cell surface.  Here
the diffusion coefficient $D\equiv c\Lambda/(\chi\rho)$.
We implement this condition by the following procedure
\citep{Turner(2004)}:
First, we assume that the requirement holds exactly at the previous step,
\begin{equation}
-D_{i,j,k+1/2}^{n-1}\frac{E_{i,j,k+1}^{n-1}-E_{i,j,k}^{n-1}}{\Delta z}=
-D_{i,j,k-1/2}^{n-1}\frac{E_{i,j,k}^{n-1}-E_{i,j,k-1}^{n-1}}{\Delta z},
\end{equation}
where $n$, $i$, $j$, and $k$ denote respectively the time step number
and the grid indexes in the $x$-, $y$-, and $z$-directions. 
Hereafter we omit the inactive indices $i$ and $j$ for clarity.
Then, we compute the ratio of radiation energy in the 
ghost cell (index $k+1$) to that in the last real cell (index $k$)
from the above equation,
\begin{equation}
r\equiv\frac{E_{k+1}^{n-1}}{E_{k}^{n-1}}=1+\frac{D_{k-1/2}^{n-1}}{D_{k+1/2}^{n-1}}\left(1-\frac{E_{k-1}^{n-1}}{E_{k}^{n-1}}\right).
\end{equation}
Finally, assuming that the ratio $r$ changes slowly over time, we obtain 
the boundary condition for the radiation energy in the current step,
\begin{equation}
E_{k+1}^{n}=rE_{k}^{n}.
\end{equation}
Therefore, the requirement holds only approximately in the current 
step.  However, this boundary condition is stable and 
also easy to implement in the numerical code.

\clearpage


\end{document}